\documentclass[letterpaper,twocolumn,10pt]{article}
\usepackage{zhanggroup}
\usepackage{latexsym}
\usepackage{tikz}
\usepackage{amsmath}
\usepackage{hyperref}
\usepackage{amsmath,amssymb,amsfonts}
\usepackage{multirow}
\usepackage{color}
\usepackage{xcolor}
\usepackage{url}
\usepackage{caption}
\usepackage{subcaption}
\usepackage{float}
\usepackage{balance}
\usepackage{graphicx}
\usepackage{textcomp}
\usepackage{booktabs}
\usepackage{colortbl}
\usepackage{pifont}
\usepackage{graphicx}
\usepackage[title]{appendix}
\usepackage[ruled,linesnumbered]{algorithm2e}  
\usepackage{algorithmic}
\usepackage{booktabs}
\usepackage{makecell}
\usepackage{tcolorbox}
\newcommand{\mypara}[1]{\noindent\textbf{#1.}\xspace}
\usepackage{filecontents}
\usepackage[absolute]{textpos}
\usepackage{caption}
\usepackage{subcaption}
\usepackage{float}
\usepackage{authblk}
\usepackage{placeins}
\usepackage{multirow}
\usepackage{booktabs}
\usepackage{algorithmic}
\usepackage{amsmath,amsfonts}
\usepackage{textcomp}
\usepackage{xcolor}
\usepackage{xspace}
\usepackage{xurl}
\usepackage{amsmath}
\usepackage{tikz}
\usepackage{filecontents}
\usepackage{booktabs} 

\begin{document}

\date{}

\title{\bf BudgetLeak: Membership Inference Attacks on RAG Systems \\ via the Generation Budget Side Channel}

\author[1]{Hao Li}
\author[1]{Jiajun He}
\author[1]{Guangshuo Wang}
\author[1]{Dengguo Feng}
\author[2,\dag]{Zheng Li}
\author[1,\dag]{Min Zhang}

\affil[1]{Institute of Software, Chinese Academy of Sciences}
\affil[2]{Shandong University}

\maketitle

\begingroup
\renewcommand{\thefootnote}{\dag}
\footnotetext{Corresponding authors.}
\endgroup

\begin{abstract}
Retrieval-Augmented Generation (RAG) enhances large language models by integrating external knowledge, but reliance on proprietary or sensitive corpora poses various data risks, including privacy leakage and unauthorized data usage.
Membership inference attacks (MIAs) are a common technique to assess such risks, yet existing approaches underperform in RAG due to black-box constraints and the absence of strong membership signals.

In this paper, we identify a previously unexplored side channel in RAG systems: the generation budget, which controls the maximum number of tokens allowed in a generated response. 
Varying this budget reveals observable behavioral patterns between member and non-member queries, as members gain quality more rapidly with larger budgets.
Building on this insight, we propose BudgetLeak, a novel membership inference attack that probes responses under different budgets and analyzes metric evolution via sequence modeling or clustering.
Extensive experiments across four datasets, three LLM generators, and two retrievers demonstrate that BudgetLeak consistently outperforms existing baselines, while maintaining high efficiency and practical viability. 
Our findings reveal a previously overlooked data risk in RAG systems and highlight the need for new defenses.
\end{abstract}

\section{Introduction}

With the rapid expansion of training corpora and model sizes, large language models (LLMs) such as GPT-4~\cite{achiam2023gpt}, LLaMA~\cite{touvron2023llama}, Mistral~\cite{jiang2023mistral7b}, and ChatGLM~\cite{zengglm} have demonstrated remarkable generative capabilities, significantly impacting both professional domains and everyday interactions. Despite these advances, LLMs continue to face fundamental challenges, including hallucinations~\cite{shuster2021retrieval,yao2023llm} and outdated knowledge~\cite{gao2023retrieval,kandpal2023large}, which raise serious concerns about their reliability in real-world applications. A widely adopted approach to address these issues is Retrieval-Augmented Generation (RAG)~\cite{lewis2020retrieval,gao2023retrieval, fan2024survey}, which improves response quality by incorporating up-to-date and contextually relevant external information into the generation process.




At the same time, the dependence of RAG systems on external databases, particularly those containing sensitive information, introduces significant privacy and security risks.
In the healthcare domain, systems such as \textit{MedPlan} may enrich with patient-specific information retrieved from external sources, enhancing the quality of clinical assessments and personalized recommendations. Such integration, however, raises concerns about unauthorized access to highly sensitive patient data~\cite{hsu2025medplan}.
A similar issue arises in the financial sector, where RAG systems employed for financial intent detection like \textit{BAI-Fintent} often rely on key transaction records, which—if improperly secured—could expose confidential financial information~\cite{Srivastava2024LendingAE, wang2025finsage}.
In the cryptography domain, tools like \textit{CRYPTOSCOPE} search for vulnerability cases stored in an external database to support cryptographic logic vulnerability detection. 
This practice increases the likelihood of security preferences being exposed~\cite{li2025cryptoscope}.
These examples underscore the importance of carefully examining how sensitive data is ingested and protected in RAG-based applications.



\begin{figure}[!t]
\centering
\begin{subfigure}{0.49\columnwidth}
\includegraphics[width=\columnwidth]{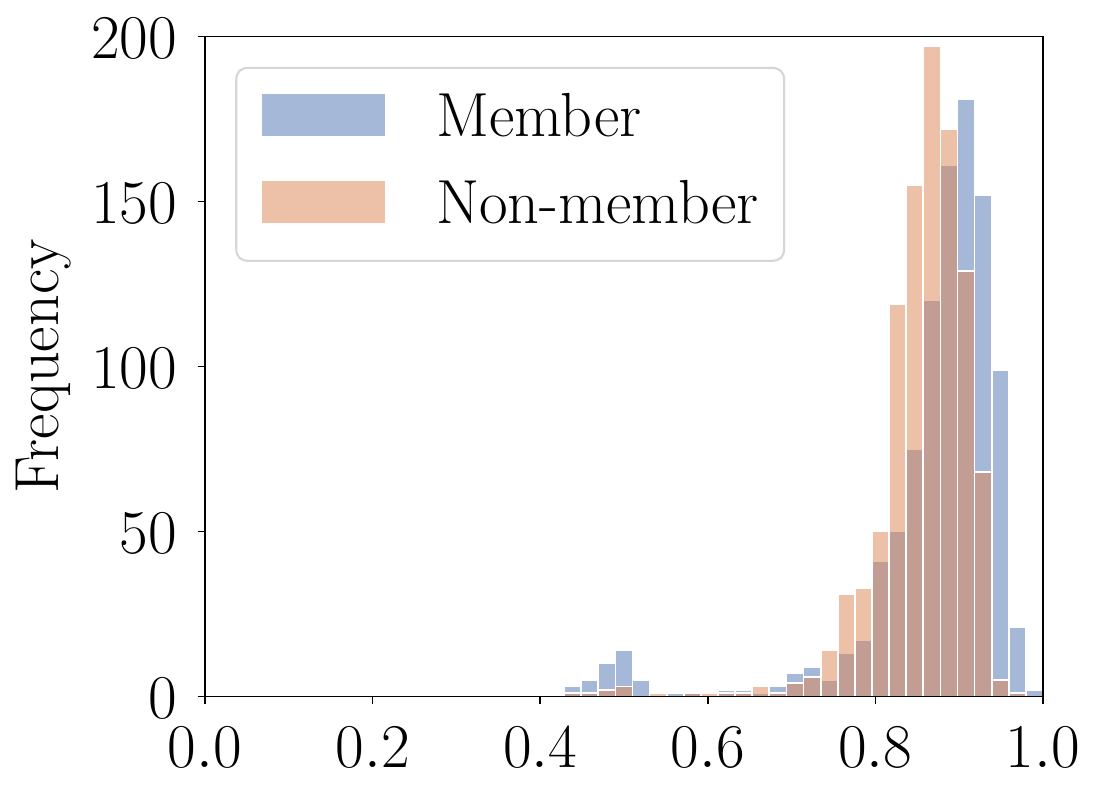}
\caption{Similarity Scores}
\label{fig:similarity}
\end{subfigure}
\begin{subfigure}{0.49\columnwidth}
\includegraphics[width=\columnwidth]{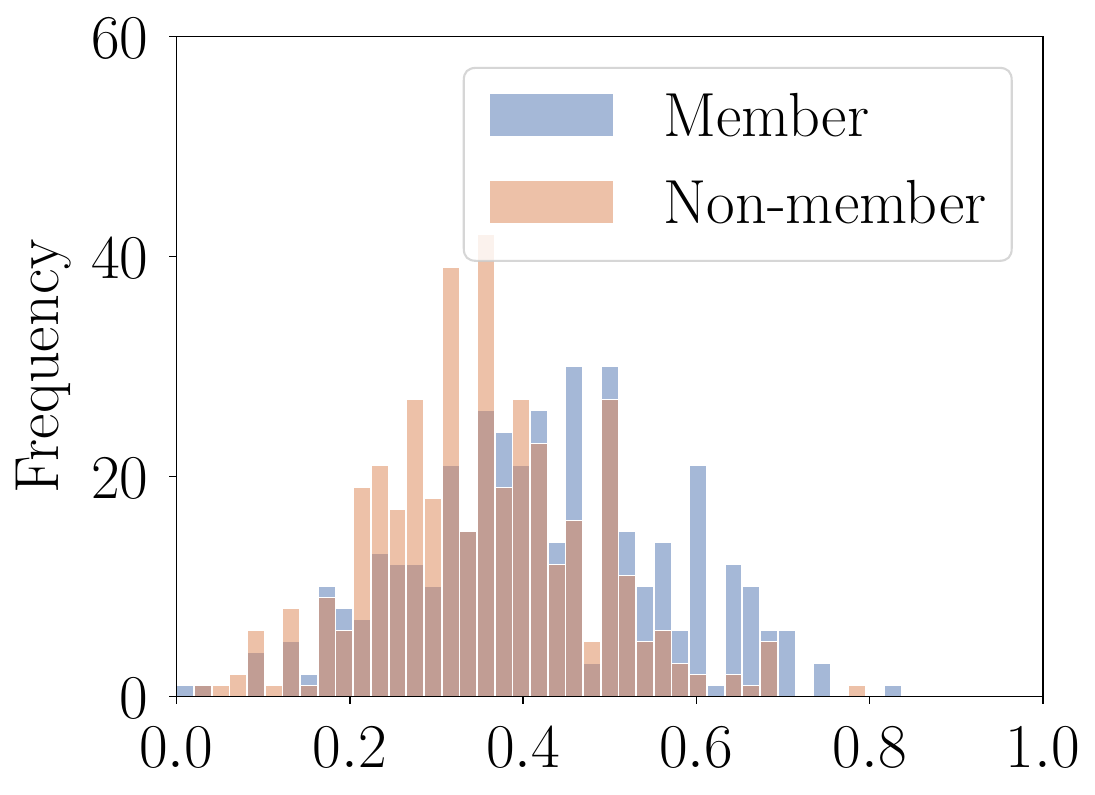}
\caption{Response Accuracy}
\label{fig:responseAccuracy}
\end{subfigure}
\caption{Distribution of similarity scores and response accuracy on RAG with LLaMA over HealthCareMagic-100k.}
\label{distributions}
\end{figure}


To address this challenge, we investigate membership inference attacks (MIAs), a well-established class of techniques for evaluating privacy risks and unauthorized data exposure.
MIAs assess whether a given sample was included in a model’s training set by analyzing the model’s prediction behavior.
In RAG systems, however, the goal shifts to identifying whether a data sample resides in the system’s knowledge base.
Several methods have been proposed for this purpose.
Li et al.~\cite{li2025generating} divided a target sample into two segments, submitted the first segment to the RAG system, and measured the similarity between the generated continuation and the second segment as a membership indicator. 
Liu et al.~\cite{liu2025mask} masked key tokens in the target sample and queried the RAG system to fill in the blanks, using fill-in accuracy as the membership score. 
Naseh et al.~\cite{naseh2025riddle} converted the target sample into a set of Yes/No questions and evaluated the RAG system’s responses, with accuracy serving as the membership signal.
These approaches compensate for the lack of conventional numerical signals (e.g., loss, confidence, entropy) in RAG outputs. 
In general, member samples yield higher similarity, fill-in accuracy, and response accuracy, reflecting better overall output quality.
Unfortunately, these signals are insufficient for reliable membership inference. 
As shown in \autoref{distributions} (see more in Appendix \autoref{fig:fill-inAccuracy}), members and non-members exhibit substantial overlap across these metrics.
This ambiguity stems from the fact that RAG output quality is influenced by multiple factors beyond membership.
First, responses quality depends not only on whether the input appears in the external knowledge base, but also on whether the internal language model encodes semantically related information. 
Bi et al.~\cite{bi2025parameters} show that the balance between internal and external knowledge can be tuned during generation, but such control is infeasible in real-world black-box scenarios.
Second, response quality varies with sample diversity, a well-known challenge in membership inference~\cite{watson2021importance,carlini2022membership,sablayrolles2019white,he2024difficulty}.
Wang et al.~\cite{wang2025rag} address this by assuming that the adversary has access to all target samples and apply a one-time calibration using shared in/out RAGs. 
While effective, this reduces the MIA task to ranking samples within a fixed dataset, which is unrealistic in practice. 
Therefore, extracting reliable membership signals from RAG systems without such assumptions remains technically challenging.

To fill this gap, we introduce a previously unexplored side-channel for membership inference: the \textit{generation budget}, defined as the maximum token limit for generated responses.
To our knowledge, this is the first study to show that this parameter systematically leaks membership information in RAG systems. 
When the generation budget is small, responses are compressed, often omitting details and reducing similarity to the ground-truth answer. 
With a large budget, the system produces more informative and specific content, yielding outputs that better align semantically with the ground truth.
This effect differs between member and non-member queries. 
For member queries, where the ground-truth answer is typically included in the retrieved context, larger budgets enable the generator to reconstruct detailed responses more effectively, leading to sharper improvements in similarity or fidelity metrics. 
Non-member queries show weaker gains under the same conditions.
This divergence in response patterns constitutes a novel and practical side channel for membership inference attacks.

Based on this observation, we present BudgetLeak, a novel side-channel membership inference attack that exploits the generation budget as a leakage signal in RAG systems.
Specifically, BudgetLeak leverages generation budget sensitivity to distinguish members from non-members.
The adversary repeatedly queries the RAG system with a target sample under varying the generation budget (e.g., 10, 20, 30 tokens) until it exceeds the ground-truth answer length.
For each response, the adversary computes its semantic similarity or other fidelity metrics relative to the ground-truth answer. 
These metric values are then ordered by the generation budget to form sequences, which are aggregated into a unified multi-metric representation.
This representation captures how metrics evolve with increasing budgets, including growth patterns, fluctuations, and cross-metric correlations. 
The adversary inputs this representation into a sequence attack model, such as a Recurrent Neural Network (RNN), to learn the temporal and relational patterns that distinguish members from non-members.
We also propose an unsupervised variant that removes the need for model training. Here, the adversary directly clusters the feature vectors derived from metric sequences, with the cluster showing higher similarity identified as the member group.

We conduct extensive experiments on four benchmark datasets across multiple RAG configurations, combining two retrievers with three LLMs. 
BudgetLeak consistently outperforms baseline methods in nearly all scenarios. 
For example, with LLaMA as the generator and MiniLM as the retriever, BudgetLeak achieves 0.982 accuracy on the HealthCareMagic-100k dataset, compared to a maximum of 0.761 by the baselines. 
Furthermore, we investigate factors affecting attack performance and evaluate the robustness under various defense mechanisms. 
The results confirm that BudgetLeak achieves the best attack performance. 

In summary, our contributions are as follows:
\begin{itemize}
    \item 
    We identify an unexplored side channel in RAG systems, the \textit{generation budget}, defined as the maximum token limit for generated responses. We reveal distinct patterns between member and non-member queries and further analyze the root causes.
    \item We propose BudgetLeak, a side-channel membership inference attack that exploits generation budget sensitivity to distinguish members from non-members. BudgetLeak operates under two adversarial settings: (i) a partial-knowledge adversary, who can create a shadow RAG using knowledge of both the data distribution and the RAG architecture, and train a sequential attack model on multi-metric sequences; and (ii) a zero-knowledge adversary, who lacks any knowledge of the RAG and relies on unsupervised clustering of feature vectors from these sequences.
    
    
    \item We evaluate BudgetLeak through extensive experiments, showing that it consistently outperforms all baselines. We also analyze key factors that affect its effectiveness, and confirm that BudgetLeak consistently achieves the best attack performance.
\end{itemize}

\section{Preliminaries and Related Works}
 
\subsection{Retrieval-Augmented Generation}

Retrieval-Augmented Generation (RAG), first introduced by Lewis et al.\cite{lewis2020retrieval}, integrates retrieval mechanisms with generative models by dynamically incorporating external knowledge as contextual input. 
This approach reduces reliance on static parameters and has inspired extensive follow-up work to improve retrieval accuracy and output relevance\cite{rag1,rag2,rag3,jiang2023active}.

Recent research has extended RAG systems with more expressive knowledge representations. One direction is Graph-RAG, which constructs knowledge graphs or semantic networks to capture inter-entity relationships, enhancing semantic coverage and contextual coherence~\cite{han2024retrieval,peng2024graph}. Another line of work focuses on multimodal RAG, which integrates information from multiple modalities such as text, images, and video to better reflect the heterogeneous nature of real-world knowledge~\cite{mmed1,mmed2}.

While these developments aim to improve generation quality, the data risks of traditional RAG systems, such as privacy leakage and unauthorized data usage, remain underexplored. Motivated by this gap, our work focuses on the canonical RAG pipeline, which consists of two components: a retriever and a generator.




\mypara{Retriever} The retriever $R$ is designed to search for $k$ relevant results $X = \{x_1, x_2, ..., x_k\}$ from an external knowledge base $D$ given an input query $q$, as formalized in Equation~\ref{eq1}. 


\begin{equation}
    \label{eq1}
    X=R(q,D).
\end{equation}

\mypara{Generator} The generator $G$ takes the retrieved results $X$ as contextual information for the input query $q$ and guides a large language model $M$ to generate an answer token by token. The generation of the next token $t_l$ is defined in Equation~\ref{eq2}. The final answer $a$ consists of the generated tokens $t_{1:n}$, that is, $a = t_{1:n}$.


\begin{equation}
    \label{eq2}
    t_l=G(q,X,t_{1:l-1},M).
\end{equation}

\subsection{Metrics for RAG}
Here, we introduce several widely used metrics for evaluating the quality of text generated by RAG systems.

\mypara{Semantic Similarity}Semantic similarity is a key metric for assessing the quality of RAG outputs~\cite{csakar2025maximizing, xia2015learning}. While various methods exist for measuring semantic similarity, following ~\cite{li2025generating}, we adopt cosine similarity as a standard approach in this work. It computes similarity by encoding the generated and original texts into embedding vectors using the same model, and then measuring the cosine of the angle between them. Higher scores indicate stronger semantic alignment and better generation quality.

\mypara{ROUGE}ROUGE (Recall-Oriented Understudy for Gisting Evaluation)~\cite{lin2004rouge, he2025context} is a widely used metric for evaluating the quality of generated text. In the context of RAG systems, it measures lexical similarity by computing n-gram overlap between generated responses and ground-truth references. Common variants include ROUGE-N, which calculates n-gram overlap, and ROUGE-L, which captures the longest common subsequence. Higher ROUGE scores indicate stronger lexical recall concerning the reference.

\mypara{BLEU}BLEU (Bilingual Evaluation Understudy)\cite{papineni2002bleu} is another widely used metric, originally proposed for machine translation. It computes n-gram precision between generated outputs and reference texts. BLEU has also been adopted in RAG evaluation settings\cite{he2025context} to quantify lexical overlap. Unlike ROUGE, which emphasizes recall, BLEU focuses on precision. Higher BLEU scores indicate greater lexical precision and thus better alignment with the reference.

\mypara{Edit Distance} Edit distance, also known as Levenshtein distance~\cite{yujian2007normalized}, measures the minimum number of edit operations, including insertions, deletions, and substitutions, required to transform one string into another. A smaller edit distance indicates greater character-level similarity between generated responses and ground-truth references, corresponding to higher generation quality.

\subsection{Membership Inference Attacks}
Membership inference attacks (MIAs) aim to determine whether a given sample was used in a model’s training set. Since the seminal work by Shokri et al.~\cite{shokri2017membership}, MIAs have been extensively studied across various settings, including white-box~\cite{sablayrolles2019white}, black-box~\cite{yeom2018privacy,salem2018ml}, label-only~\cite{li2021membership,lienhanced}, and federated learning~\cite{nasr2019comprehensive,chang2024efficient}.

Early approaches, such as those by Shokri et al.~\cite{shokri2017membership} and Salem et al.~\cite{salem2018ml}, introduced the shadow model framework, where multiple shadow models simulate the target model’s behavior, and an attack classifier is trained on their outputs. Later, Song et al.~\cite{song2021systematic} proposed a more direct threshold-based method, which compares metrics (e.g., loss, entropy) derived from model outputs against predefined thresholds. Carlini et al.~\cite{carlini2022membership} further improved this by introducing difficulty calibration, adjusting thresholds based on the learning difficulty of each sample. This idea was expanded in subsequent studies~\cite{ye2022enhanced,watson2021importance,he2024difficulty}. More recently, Li et al.~\cite{li2024seqmia} demonstrated that membership signals can be effectively extracted during training. With the rise of LLMs, MIAs have also been explored in this context. The Min-K\% Prob method~\cite{shi2023detecting} assumes that tokens with higher predicted probabilities are more likely to be from training data. Zhang et al.~\cite{zhang2024pretraining} proposed a divergence-based calibration method that compares the token probability distribution with the token frequency distribution to derive detection scores.

In contrast to conventional LLMs, RAG incorporates external information from a knowledge base during generation, without requiring fine-tuning. The MIA goal in this setting is to determine whether a given sample $x$ is present in the knowledge base $D$. To achieve this, adversaries craft targeted prompts and analyze the RAG’s responses. For example, Anderson et al.~\cite{anderson2024my} directly queried the RAG system about the presence of $x$ in the knowledge base, though this was shown to be easily mitigated using defense prompts~\cite{naseh2025riddle}. Further, Li et al.~\cite{li2025generating} truncated the target sample and inferred membership by measuring the cosine similarity between the RAG-generated continuation and the omitted portion. Wang et al.~\cite{wang2025rag} extended this approach by applying a likelihood ratio test to further improve attack accuracy. Liu et al.~\cite{liu2025mask} designed a mask-based attack, in which specific tokens are masked and the accuracy of their recovery is used for inference. Naseh et al.~\cite{naseh2025riddle} constructed a set of Yes/No questions from the target sample and queried the RAG system for answers, using response accuracy as a membership signal.

\section{Attack Methodology}

\subsection{Threat Model}
We study membership inference attacks against RAG systems under black-box access. We consider two adversarial settings: a \emph{partial-knowledge adversary}, which aligns with prior work, and a stricter \emph{zero-knowledge adversary}, which better reflects real-world deployments.

\mypara{Adversarial Scenario}
The adversary interacts with the RAG system only through its public interface. By submitting queries and observing textual outputs, the adversary aims to infer whether a target sample belongs to the private knowledge base $D$.

\mypara{Adversary’s Goal}
The adversary's objective is to determine membership of specific data records in $D$ without direct access to the retriever $R$, the generator $G$, or intermediate computations. 
This can violate data confidentiality. 
Besides, data owners can also leverage this technique to protect their data copyright when data is used without permission.

\mypara{Adversary’s Capabilities}
We distinguish two levels of adversarial knowledge:

\begin{itemize}
\item \textbf{Partial-knowledge adversary.} This model reflects assumptions made in prior studies~\cite{li2025generating, liu2025mask, wang2025rag, naseh2025riddle}.
\begin{itemize}
\item \textit{Black-box access:} The adversary can query the system and observe outputs only.
\item \textit{Knowledge of data distribution:} The adversary knows the distribution underlying $D$ and can sample from it, but cannot access $D$ directly.
\item \textit{Knowledge of architecture:} The adversary knows the high-level structure of the RAG system, e.g., retriever and generator types.
\end{itemize}

\item \textbf{Zero-knowledge adversary.} This stricter model captures more realistic deployment scenarios.  
\begin{itemize}  
    \item \textit{Black-box access only:} Same as above.  
    \item \textit{No prior knowledge:} The adversary lacks any information about the data distribution or the RAG system’s architecture.  
\end{itemize}  

\end{itemize}

The zero-knowledge adversary represents a practical yet underexplored threat model in current literature.

\begin{figure}[!t]
\centering
\begin{subfigure}{0.49\columnwidth}
\includegraphics[width=\columnwidth]{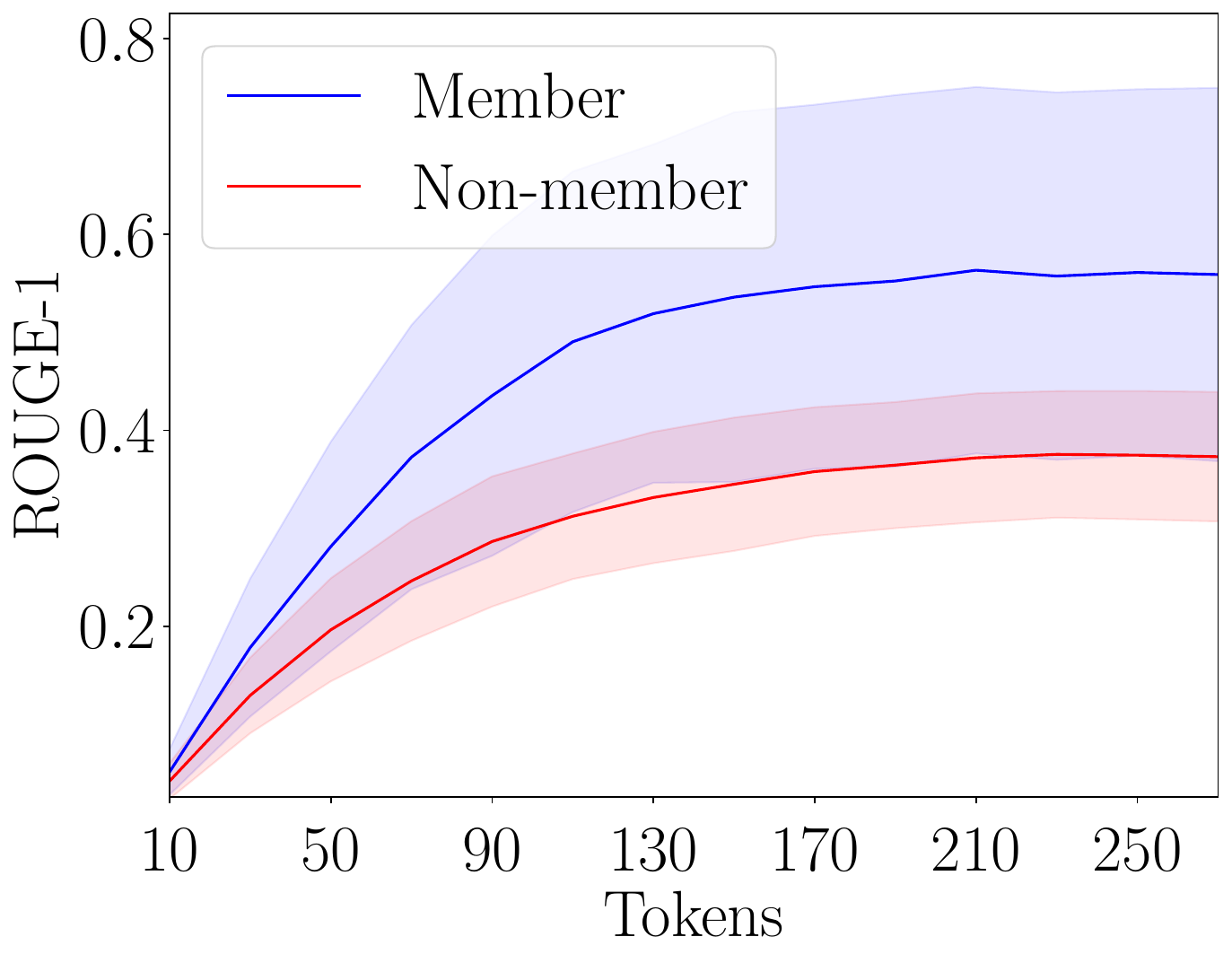}
\caption{Change Rate of ROUGE-1}
\label{fig:rate_of_change}
\end{subfigure}
\begin{subfigure}{0.49\columnwidth}
\includegraphics[width=\columnwidth]{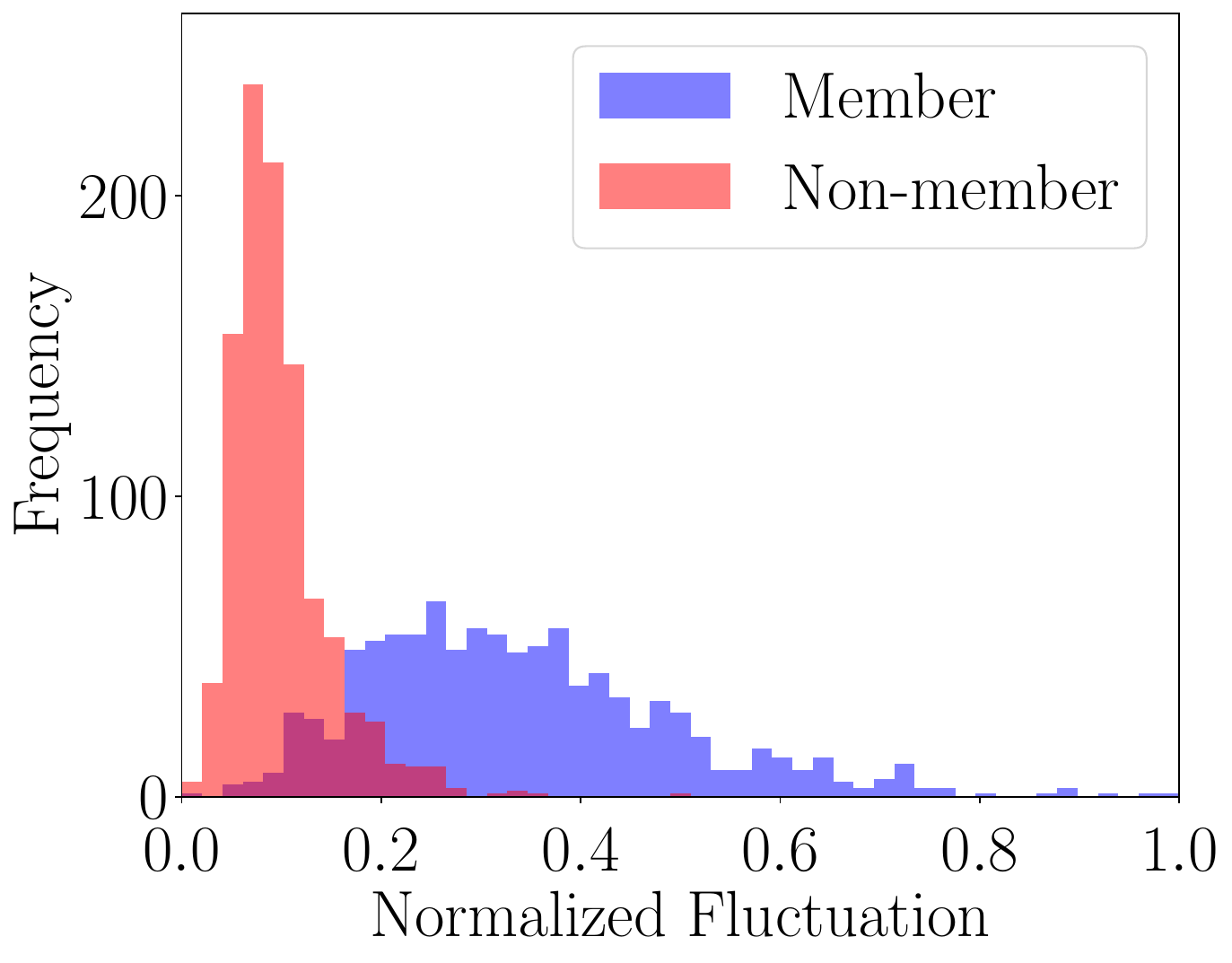}
\caption{Fluctuations of ROUGE-1}
\label{fig:fluctuation_distribution}
\end{subfigure}

\caption{Metric change rate and cumulative fluctuation distribution under varying generation budgets in RAG with LLaMA on HealthCareMagic-100k.}
\label{pattens1and2}
\end{figure}

\begin{figure}[!t]
\centering
\begin{subfigure}{0.49\columnwidth}
\includegraphics[width=\columnwidth]{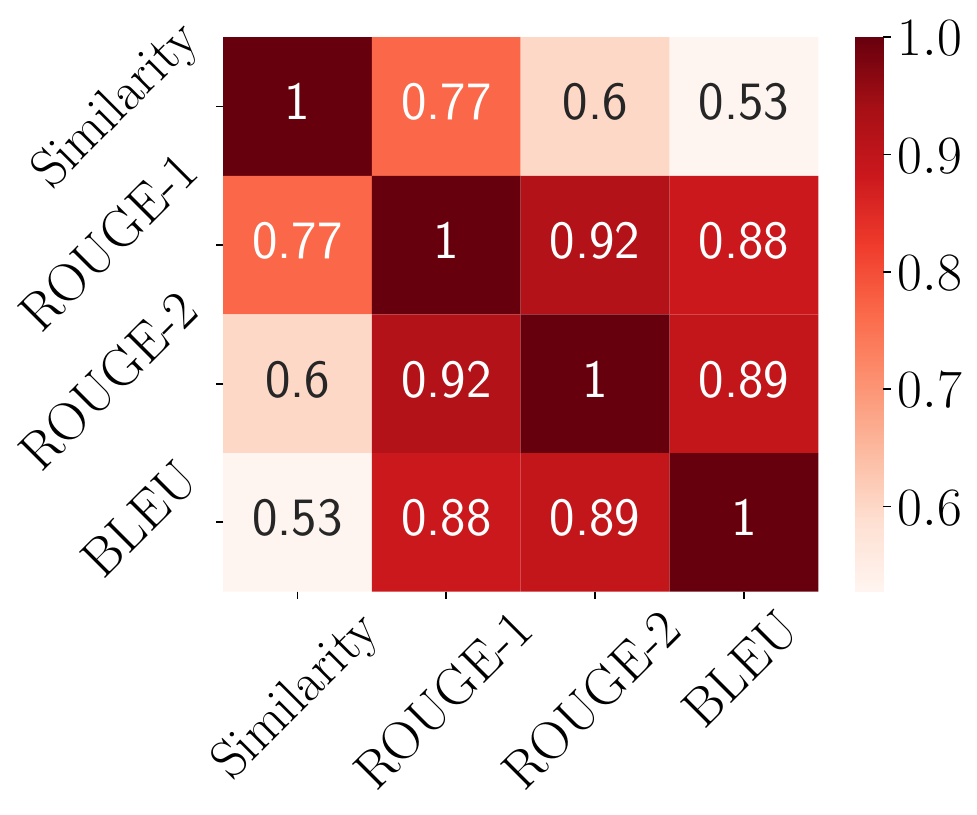}
\caption{Member}
\label{fig:correlation_Member}
\end{subfigure}
\begin{subfigure}{0.49\columnwidth}
\includegraphics[width=\columnwidth]{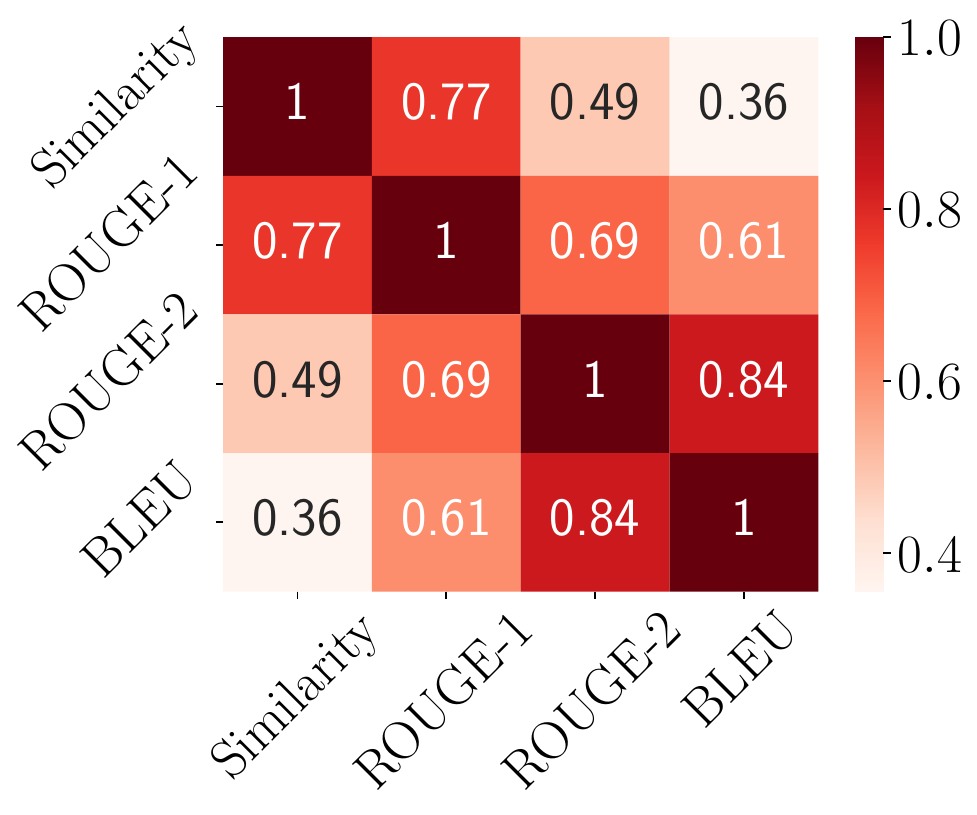}
\caption{Non-member}
\label{fig:correlation_Non-member}
\end{subfigure}

\caption{Differences in cross-metric correlations between members and non-members in RAG with LLaMA on HealthCareMagic-100k.}
\label{pattens3}
\end{figure}

\subsection{Design Intuition}\label{Design-Intuition}
As noted earlier, existing MIAs on RAG systems~\cite{li2025generating, liu2025mask, naseh2025riddle} often suffer from weak membership signals. 
This motivates us to examine alternative channels of membership leakage. We observe a strong correlation between output quality (i.e., closeness to ground truth) and generation budget (i.e., output length). 
In general, a larger generation budget enables a RAG system to produce more detailed responses, thereby improving answer quality. 
Building on this intuition, we analyzed the mechanism and identified three key patterns, namely rate-of-change, cumulative fluctuation, and cross-metric correlations.

\mypara{Rate-of-Change} We first observe that RAG systems exhibit a much faster quality improvement in response quality for member queries as the generation budget increases.
\autoref{fig:rate_of_change} shows this trend using ROUGE-1 scores over 1,000 member and 1,000 non-member queries, with the budget ranging from 10 to 270 tokens.
This discrepancy stems from the RAG system’s ability to retrieve ground-truth content for member queries, enabling both semantic and lexical alignment. 
With a larger budget, the system efficiently approximates the correct answer. 
In contrast, non-member queries lack such support and rely solely on the generator’s parametric knowledge, resulting in slower and often lower-quality improvements.
We define this varying rate of quality improvement as a rate-of-change signal.

\mypara{Cumulative Fluctuation} Next, we analyze how the output quality fluctuates with increasing generation budget, and whether this behavior differs between member and non-member samples. \autoref{fig:fluctuation_distribution} shows the distribution of cumulative quality fluctuation, measured as the sum of absolute differences in ROUGE-1 scores between adjacent generation budgets, for 1,000 member and 1,000 non-member queries. 
A clear divergence emerges: member samples exhibit greater fluctuation.
This occurs because, as the budget grows, member queries leverage ground-truth content from the knowledge base, producing sharper quality gains.
In contrast, non-member queries lack such support, resulting in smaller variations.
We refer to this divergence as a cumulative fluctuation signal.

\mypara{Cross-metric Correlations} Furthermore, we explore whether different quality metrics—such as semantic similarity and lexical measures like ROUGE-1 and BLEU—exhibit consistent trends as the generation budget increases. \autoref{pattens3} shows the correlation of these metrics across varying budgets. 
We observe that member queries tend to exhibit stronger cross-metric correlations. 
For instance, the correlation between ROUGE-1 and BLEU is 0.88 for member queries, compared to 0.61 for non-member queries.
This occurs because RAG responses for member queries tend to align with the ground-truth answer both semantically and lexically. 
In contrast, non-member responses may be semantically close to the ground-truth but use different wording, resulting in lower lexical similarity.
We refer to this distinction as a correlation signal between member and non-member samples.

\mypara{Attack Strategy} At a high level, our attack strategy queries the RAG system multiple times for each target data sample, varying the generation budget each time.
Then we record quality metrics at each budget to form a multi-metric sequence that captures how output quality evolves. 
We finally distinguish members from non-members by leveraging patterns in quality dynamics and cross-metric correlations.

\begin{figure*}[!h]
    \centering
    \includegraphics[width=0.95\linewidth]{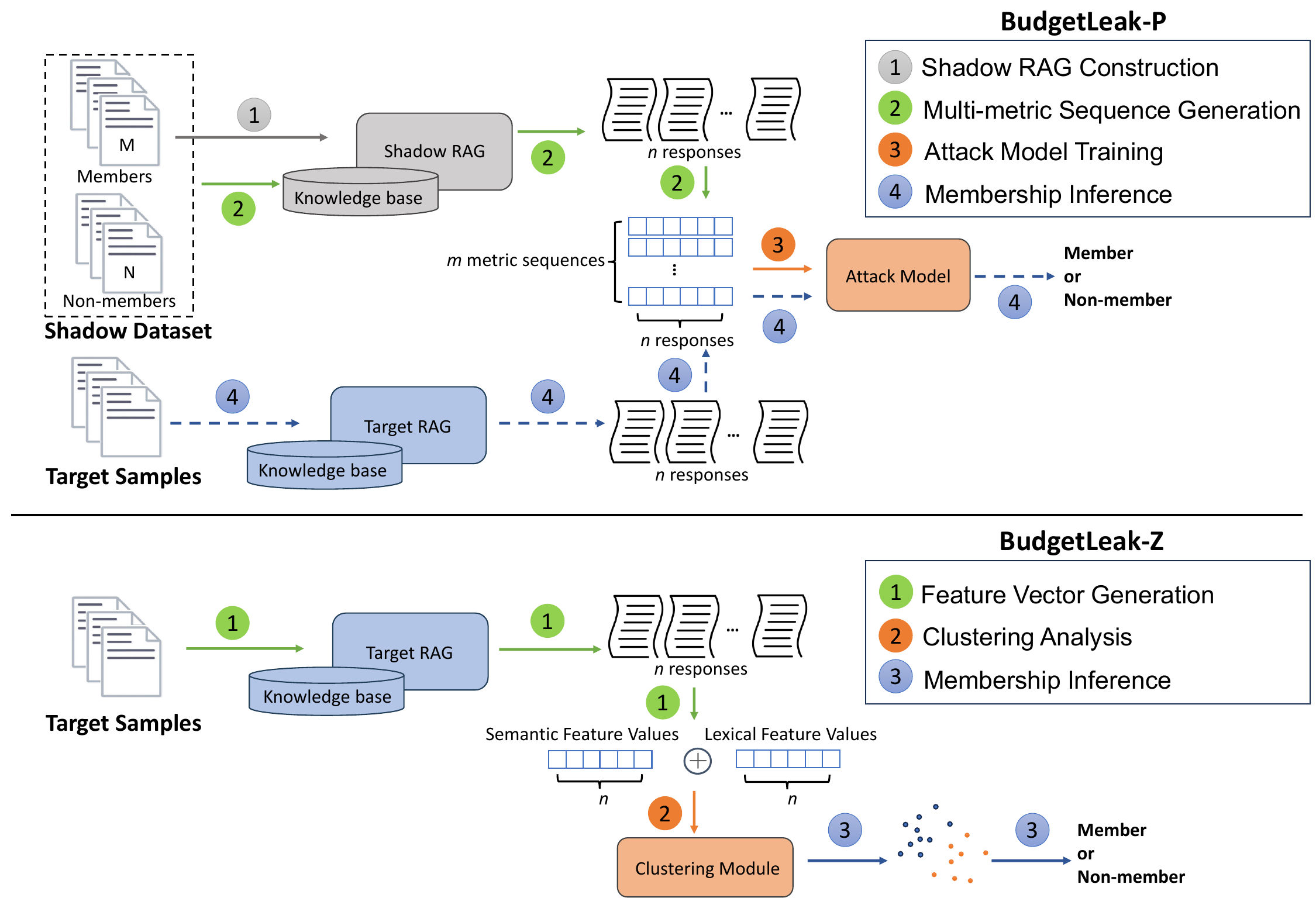}
    \caption{Overview of BudgetLeak.
    }
    \label{Overview-of-BudgetLeak}
\end{figure*}

\subsection{Partial-knowledge Adversary}
Based on the above, we introduce BudgetLeak, a novel membership inference attack against RAG systems that exploits generation budget as a side channel. 
We refer to BudgetLeak under the partial-knowledge adversary as BudgetLeak-P.
As illustrated in \autoref{Overview-of-BudgetLeak}, the attack involves four stages: shadow RAG construction, multi-metric sequence generation, attack model training, and membership inference.

\mypara{Shadow RAG Construction} Under the partial-knowledge setting, the adversary holds a shadow dataset that is i.i.d. with the target RAG system’s knowledge base. This dataset is split into two parts: one to build a shadow RAG (member samples) and the other to serve as non-member samples.
Given that the adversary knows the generator and retriever of the target RAG system, the shadow RAG is built with the same components and an i.i.d. knowledge base.
Consequently, the shadow RAG reproduces the target’s behavior, yielding high-quality outputs for member queries and lower-quality outputs for non-member queries, as measured by semantic similarity or related metrics.

\mypara{Multi-metric Sequence Generation} The adversary builds a multi-metric sequence for each member and non-member sample of the shadow RAG system using the generation budget side channel.
Each sample is queried with $n$ different budgets, producing $n$ responses. 
For each response, a similarity score to the ground-truth answer is computed, and the scores are ordered according to increasing generation budgets, forming a similarity sequence.
Additional sequences are computed using metrics such as ROUGE-1 and BLEU. 
Combining $m$ such sequences yields an $m \times n$ matrix, called the multi-metric sequence, which captures how response quality evolves across metrics as the generation budget increases.

\mypara{Attack Model Training} Because the shadow RAG is fully controlled by the adversary, membership labels can be accurately assigned to all sequences from the previous stage, yielding a labeled dataset of multi-metric sequences. 
This dataset is then used to train a binary classifier for membership inference. 
The three membership-related patterns identified in \autoref{Design-Intuition} are sequential in nature, motivating the use of sequence models such as RNNs, attention-based RNNs, transformers, or recent variants like Mamba, which are well-suited to capture temporal dependencies. 
We evaluate several sequence-based models in \autoref{Ablation-Study} and find that the attention-based RNN performs best for our task. 
Its parameters are optimized by minimizing cross-entropy loss.

\mypara{Membership Inference} Once trained, the attack model infers membership status for target samples. The adversary follows the same procedure from the multi-metric sequence generation stage, querying the target RAG system multiple times under varying generation budgets. 
The responses generate a multi-metric sequence for the target sample. 
The attack model processes this sequence and predicts whether the sample belongs to the target system's knowledge base.

\subsection{Zero-knowledge Adversary}
We now present the zero-knowledge adversary, where the adversary lacks knowledge of the target RAG system's internal configurations and does not have access to a shadow dataset similar to the target knowledge base. 
This realistic scenario remains unexplored in prior MIA studies on RAG systems~\cite{li2025generating, liu2025mask, naseh2025riddle, wang2025rag}. 
To address this gap, we propose BudgetLeak-Z, a zero-knowledge MIA method using clustering techniques. 
As illustrated in \autoref{Overview-of-BudgetLeak}, the method comprises three stages: feature vector generation, clustering analysis, and membership inference.


\mypara{Feature Vector Generation} Similar to the Multi-metric Sequence Generation stage in BudgetLeak-P, the adversary leverages the generation budget side channel to obtain sequences of similarity scores, ROUGE-1, or other metric values for each sample. To satisfy the input requirements of clustering algorithms, these sequences are not aligned and assembled into a matrix but are instead directly concatenated into a single feature vector. To mitigate the impact of high dimensionality, one semantic and one lexical metric sequence are selected and combined to capture how the sample’s response quality evolves with increasing generation budgets.

\mypara{Clustering Analysis} The adversary feeds the feature vectors of all target samples into a clustering module with two clusters, representing members and non-members. 
To identify the member cluster, the adversary compares the average quality metrics of the clusters under the same generation budget. 
For example, in the case of similarity, the cluster with the higher average score is labeled as the member cluster, since member samples generally produce higher-quality responses.

\mypara{Membership Inference} Based on the clustering results, the adversary infers membership by classifying samples in the member cluster as members and those in the other cluster as non-members. If fuzzy clustering algorithms such as Fuzzy C-Means (FCM)~\cite{bezdek1984fcm} are used in the Clustering Analysis stage, the adversary obtains membership probabilities for each sample, allowing more flexible thresholding to refine predictions. However, FCM is not mandatory—other clustering methods like K-means can also be used. Finally, for new samples not involved in the initial clustering, classification algorithms like K-Nearest Neighbors (KNN)~\cite{cunningham2021k} can be applied to infer membership without reclustering or retraining.

\section{Experimental Setup}

\subsection{Datasets}
To thoroughly evaluate the performance of our method, we conduct experiments on three question-answering (QA) datasets and one text classification dataset, covering both specialized and open-domain scenarios. \textbf{HealthCareMagic-100k}~\cite{HealthCareMagic} consists of real-world conversations between patients and doctors. \textbf{MS-MARCO}~\cite{bajaj2016ms} and \textbf{Natural Questions}~\cite{kwiatkowski2019natural} are widely used open-domain QA datasets released by Microsoft and Google, respectively. \textbf{AGNews}~\cite{zhang2015character} is a four-class news topic classification dataset collected from over 2,000 news sources. 


We then partition the datasets for both attack and evaluation purposes. First, we preprocess by removing overly short entries, retaining only samples with at least 50 tokens. Following the knowledge base size used in~\cite{liu2025mask,wang2025rag}, we randomly select 8,000 samples each to build the target RAG knowledge base $D^t$ and the shadow RAG knowledge base $D^s$, ensuring no overlap between them. To enhance evaluation stability, we use a larger number of member and non-member samples than~\cite{liu2025mask}. Specifically, we select 1,000 samples from $D^t$ and $D^s$ as member samples for the target and shadow RAG systems, denoted as $D^t_{in}$ and $D^s_{in}$, respectively. Another 1,000 samples are drawn from the remaining data $D \setminus (D^t \cup D^s)$ for each system to serve as non-member samples, forming $D^t_{out}$ and $D^s_{out}$. All subsets are mutually exclusive. 

In this setup, the adversary cannot access the target knowledge base $D^t$, including its member samples $D^t_{in}$, nor the non-member samples $D^t_{out}$. For BudgetLeak-P, the adversary has full access to the shadow dataset $D^s$, including $D^s_{in}$ and $D^s_{out}$. For BudgetLeak-Z, the adversary has no access to any shadow data.

\subsection{RAG Settings}
In our experiments, we employ three widely used open-source LLMs as RAG generators: LLaMA~\cite{touvron2023llama} (Meta-Llama-3-8B-Instruct), Mistral~\cite{jiang2023mistral7b} (Mistral-7B-Instruct-v0.2), and ChatGLM~\cite{zengglm} (GLM-4-9B-Chat). For the retriever component, we use two commonly adopted embedding models: MiniLM~\cite{all-MiniLM-L6-v2}(all-MiniLM-L6-v2) and BGE~\cite{bge-small-en-v1.5}(bge-small-en-v1.5). By default, the number of retrieved entries (top-$k$) is set to 4. We also evaluate alternative top-$k$ values in the \autoref{Ablation-Study} to assess their impact.

Similar to~\cite{li2025generating,naseh2025riddle,wang2025rag}, we set the system prompt as ``Please answer the question based on the provided context. Context: \{$context$\}. Question: \{$userInput$\}.'' This prompt is part of the RAG system’s internal configuration and cannot be modified by users or adversaries. Although slight variations may exist across different RAG systems to improve response quality, all baselines and our method use the same system prompt for consistency. 

Regarding $userInput$, which refers to the input provided by the user or adversary to the RAG system, we follow the default configurations specified by each baseline. For our method, we use the question part of each QA dataset instance as the user prompt. In the case of the AGNews dataset, we adopt the approach from~\cite{li2025generating, wang2025rag} and set $userInput$ as ``Complete this sentence \{$text$\} based on the context,'' where $text$ consists of the first 10 tokens of the news article.

\subsection{BudgetLeak Hyper-parameters}
The generation budget side channel is a core component of our approach. By default, we vary the generation budget, which refers to the maximum number of tokens that can be generated, from 10 to 270 in steps of 20. Each query is issued once under each budget setting. To further evaluate the practicality of our method, we also explore its effectiveness with fewer queries to the target RAG system, as discussed in \autoref{practialdiscussion}. In addition, we adopt an attention-based LSTM as the attack model. The number of training epochs ranges from 200 to 500, depending on the complexity of the datasets. The learning rate is initially set to 0.01 and gradually decays to 0.0005.

\subsection{Baselines}
To demonstrate the effectiveness of our proposed methods, we compare them against several recent MIA approaches targeting RAG systems, including S$^2$MIA~\cite{li2025generating}, DC-MIA~\cite{wang2025rag}, MBA~\cite{liu2025mask}, and IA~\cite{naseh2025riddle}. Among these, MBA and IA represent the current state-of-the-art.

\mypara{S$^2$MIA}S$^2$MIA~\cite{li2025generating} splits a target sample into two parts, inputs the first part into the RAG system, and computes the similarity between the generated response and the second part. This similarity score serves as the membership inference signal, as member samples generally yield higher scores than non-members. Based on this, S$^2$MIA proposes two supervised attack variants: one selects a threshold using a shadow dataset (S$^2$MIA-T), and the other trains a binary classifier on shadow data (S$^2$MIA-M). Although S$^2$MIA also explores using response perplexity as an auxiliary signal, this requires access to internal model outputs and is thus infeasible in black-box settings. Therefore, we evaluate only the similarity-based versions, S$^2$MIA-T and S$^2$MIA-M, in our experiments.

\mypara{Difficulty-Calibrated Membership Inference Attack (DC-MIA)}DC-MIA~\cite{wang2025rag} adopts a strategy akin to S$^2$MIA for measuring the closeness between RAG-generated and ground-truth answers. Its key improvement lies in considering varying query difficulty, which impacts similarity scores and membership inference. To address this, DC-MIA calibrates membership scores via a likelihood ratio test~\cite{carlini2022membership, he2024difficulty}, leveraging multiple reference RAGs to enhance robustness and inference accuracy.

\mypara{Mask-Based Membership Inference Attack (MBA)}MBA~\cite{liu2025mask} masks a fixed number of specific tokens in a target text and feeds it to the RAG system for completion. The fill-in accuracy is used as the membership signal, since RAG systems typically perform better on member samples than on non-members.

\mypara{Interrogation Attack (IA)}IA~\cite{naseh2025riddle} uses an LLM to extract a summary from the target text and generates $n$ Yes/No questions along with their ground truth answers. For each question, the summary and the question are concatenated to form a query, which is then submitted to the target RAG system. This results in $n$ queries per text. The membership inference signal is based on the accuracy of the target RAG system’s answers, which are typically higher for member samples than for non-members.

Lastly, we adopt the default settings for key hyperparameters as specified in prior works~\cite{liu2025mask,naseh2025riddle,wang2025rag}. Specifically, DC-MIA utilizes 16 reference RAGs; MBA masks 5 or 10 tokens per target sample depending on its length; and IA generates 30 Yes/No questions per sample.

\begin{figure*}[!th]
    \centering
    \includegraphics[width=0.9\linewidth]{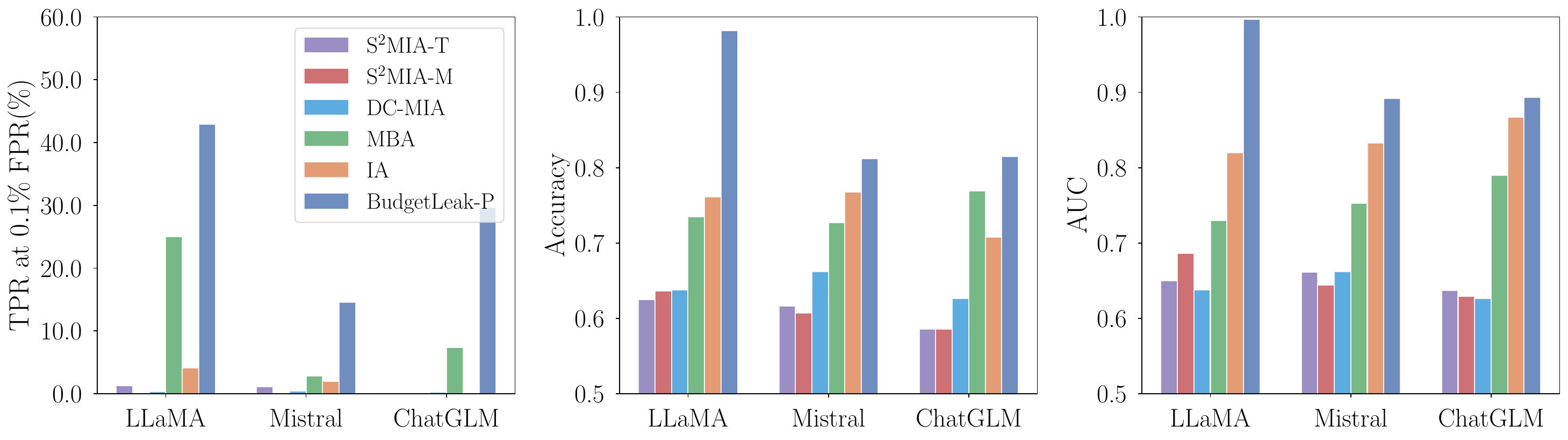}
    \caption{Performance of various attacks against RAGs on the HealthCareMagic-100k dataset (BudgetLeak-P).
    }
    \label{performance:healthcaremagic}
\end{figure*}

\begin{figure*}[!th]
    \centering
    \includegraphics[width=0.9\linewidth]
    {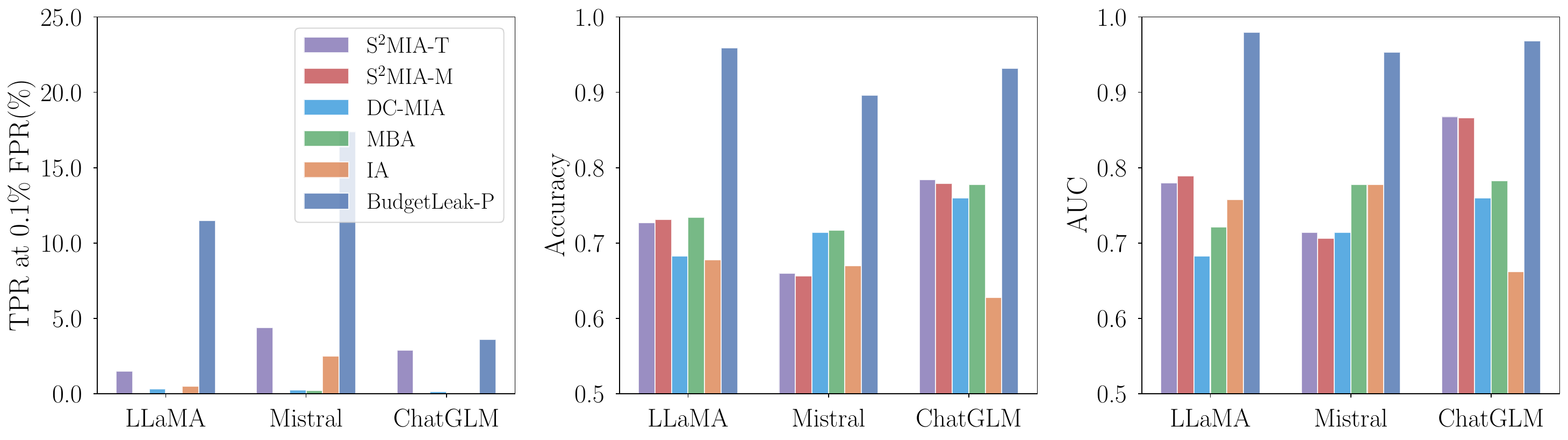}
    \caption{Performance of various attacks against RAGs on the MS-MARCO dataset (BudgetLeak-P).
    }
    \label{performance:MSMarco}
\end{figure*}

\begin{figure*}[!th]
    \centering
    \includegraphics[width=0.9\linewidth]{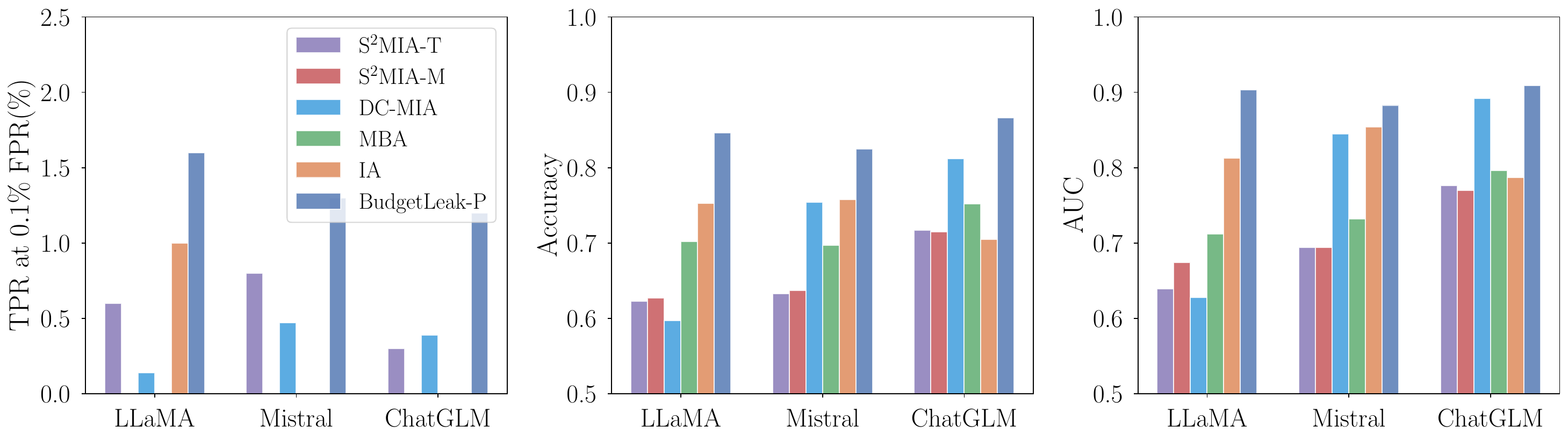}
    \caption{
    Performance of various attacks against RAGs on the Natural Questions dataset (BudgetLeak-P).
    }
    \label{performance:NQ}
\end{figure*}

\begin{figure*}[!th]
    \centering
    \includegraphics[width=0.9\linewidth]{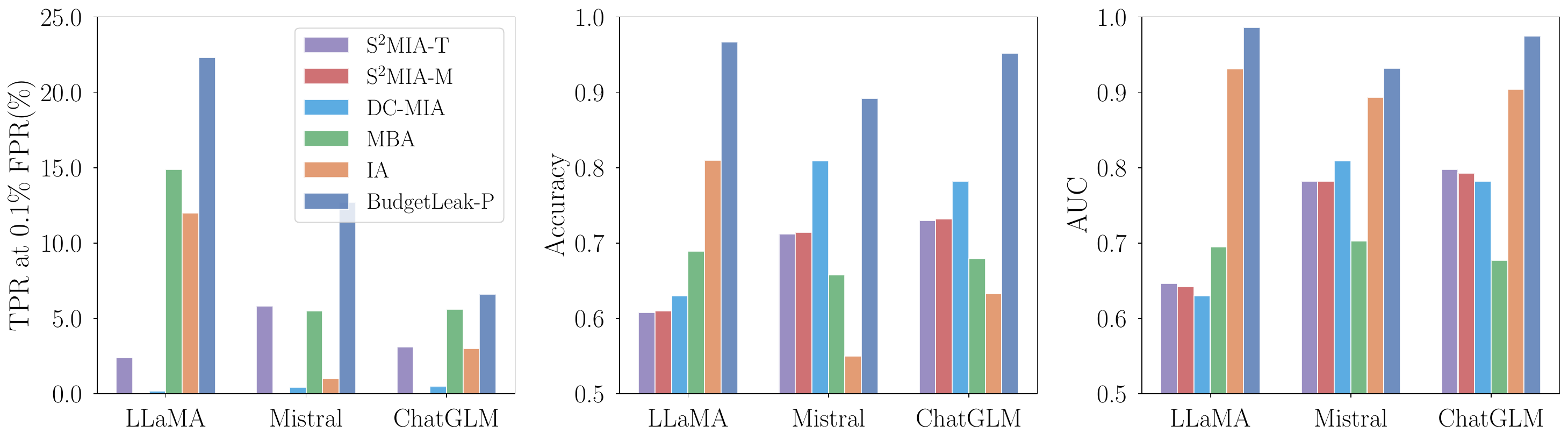}
    \caption{
    Performance of various attacks against RAGs on the AGNews dataset (BudgetLeak-P).
    }
    \label{performance:agnews}
\end{figure*}

\begin{table*}[!ht]									
\caption{Performance of various attacks against RAGs on four datasets (BudgetLeak-Z).}										
\label{cluster}
\centering
\scalebox{0.8}
{
\centering	
\small
\resizebox{\textwidth}{!}{			
\setlength{\tabcolsep}{4pt}		
\renewcommand{\arraystretch}{1}
\begin{tabular}{@{}c|cccc|cccc@{}}					
\toprule												
\multirow{2}{*}{Dataset} & \multicolumn{4}{c|}{LLaMA} & \multicolumn{4}{c}{Mistral} \\												
\cmidrule(lr){2-5} \cmidrule(lr){6-9}	
 & Method & AUC & Accuracy & \makecell{TPR at 0.1\% \\ FPR(\%)} & Method & AUC & Accuracy & \makecell{TPR at 0.1\% \\ FPR(\%)} \\	
\midrule												
\multirow{5}{*}{HealthCareMagic-100k}				
 & S$^2$MIA & 0.542 & 0.518 & 0.0 						
 & S$^2$MIA & 0.621 & 0.603 & 0.5 \\
 & DC-MIA  & 0.587 & 0.587 & 0.0							
 & DC-MIA  & 0.715 & 0.595 & \underline{1.4} \\
 & MBA & \underline{0.730} & \underline{0.735} & \underline{1.8} 
 & MBA & 0.553 & 0.553 & 0.0 \\							
 & IA  & 0.714 & 0.700 & 1.5					
 & IA  & \underline{0.809} & \textbf{0.755} & 0.0 \\
 \cline{2-9}
 \addlinespace[1ex] 
 & BudgetLeak-Z & \textbf{0.978} & \textbf{0.860} & \textbf{18.5}	
 & BudgetLeak-Z & \textbf{0.813} & \underline{0.665} & \textbf{4.8} \\ 
\midrule												
\multirow{5}{*}{Natural Questions}
 & S$^2$MIA & 0.511 & 0.516 & \textbf{0.2} 
 & S$^2$MIA & 0.643 & 0.626 & 0.1 \\ 
 & DC-MIA  & 0.521 & 0.519 & 0.0							
 & DC-MIA  & 0.626 & 0.701 & 0.2 \\
 & MBA & 0.706 & 0.702 & 0.0 
 & MBA & 0.686 & 0.661 & 0.0 \\ 
 & IA  & \underline{0.766} & \textbf{0.743} & 0.0 
 & IA  & \textbf{0.733} & \textbf{0.720} & \textbf{6.5} \\
 \cline{2-9}
 \addlinespace[1ex] 
 & BudgetLeak-Z & \textbf{0.884} & \underline{0.738} & \underline{0.1} 	
 & BudgetLeak-Z & \underline{0.689} & \underline{0.702} & \underline{0.3} \\ 
 \midrule												
\multirow{5}{*}{MS-MARCO}
 & S$^2$MIA & 0.583 & 0.553 & 0.2
 & S$^2$MIA & 0.636 & 0.614 & \textbf{0.2} \\
 & DC-MIA  & 0.551 & 0.550 & 0.0							
 & DC-MIA  & 0.645 & 0.612 & 0.0 \\
 & MBA & 0.720 & \underline{0.734} & 0.0 
 & MBA & \underline{0.773} & \underline{0.724} & 0.0\\ 
 & IA & \underline{0.723} & 0.668 & \underline{1.5}
 & IA & 0.744 & 0.710 & 0.0 \\
 \cline{2-9}
 \addlinespace[1ex] 
 & BudgetLeak-Z & \textbf{0.983} & \textbf{0.910} & \textbf{2.5} 
 & BudgetLeak-Z & \textbf{0.914} & \textbf{0.769} & \underline{0.1} \\
 \midrule												
\multirow{5}{*}{AGNews}	
 & S$^2$MIA & 0.612 & 0.610 & \underline{0.0}
 & S$^2$MIA & 0.691 & 0.644 & 0.4 \\
 & DC-MIA  & 0.641 & 0.625 & \underline{0.0}	
 & DC-MIA  & 0.799 & 0.689 & 0.3 \\
 & MBA & 0.694 & 0.693 & \underline{0.0}	
 & MBA & 0.694 & 0.657 & 0.0 \\ 
 & IA & \underline{0.910} & \underline{0.838} & \underline{0.0}	
 & IA & \underline{0.880} & \underline{0.795} & \underline{0.5} \\
  \cline{2-9}
 \addlinespace[1ex] 
 & BudgetLeak-Z & \textbf{0.978} & \textbf{0.888} & \textbf{1.4}
 & BudgetLeak-Z & \textbf{0.905} & \textbf{0.807} & \textbf{2.3} \\	
\bottomrule												
\end{tabular}
}
}												
\end{table*}	

\subsection{Evaluation Metrics}
We consider the following evaluation metrics, which are widely adopted in recent studies on membership inference attacks~\cite{carlini2022membership, li2024seqmia, he2024difficulty, li2025generating, liu2025mask, naseh2025riddle}.

\mypara{Accuracy} Accuracy is one of the most commonly used metrics for evaluating the performance of binary classification tasks. In our evaluation setup, the number of member and non-member samples is balanced, allowing this metric to reflect the model’s predictive performance on both classes. This is also referred to as balanced accuracy.

\mypara{AUC} The Area Under the Receiver Operating Characteristic Curve (AUC) measures a binary classifier’s ability to distinguish between member and non-member samples. Higher AUC indicates better discriminative performance.

\mypara{TPR at Low FPR} The True Positive Rate (TPR) at low False Positive Rate (FPR), recommended by Carlini et al.~\cite{carlini2022membership}, is widely used in recent membership inference studies~\cite{carlini2022membership, li2024seqmia, he2024difficulty, naseh2025riddle}. It reflects attack effectiveness under strict FPR constraints, highlighting practical utility. Following~\cite{carlini2022membership, li2024seqmia, he2024difficulty}, we report TPR at 0.1\% FPR.

\section{Experimental Results}

\begin{figure*}[h]
    \centering
    \begin{subfigure}[b]{0.3\linewidth}
        \centering
        \includegraphics[width=\linewidth]{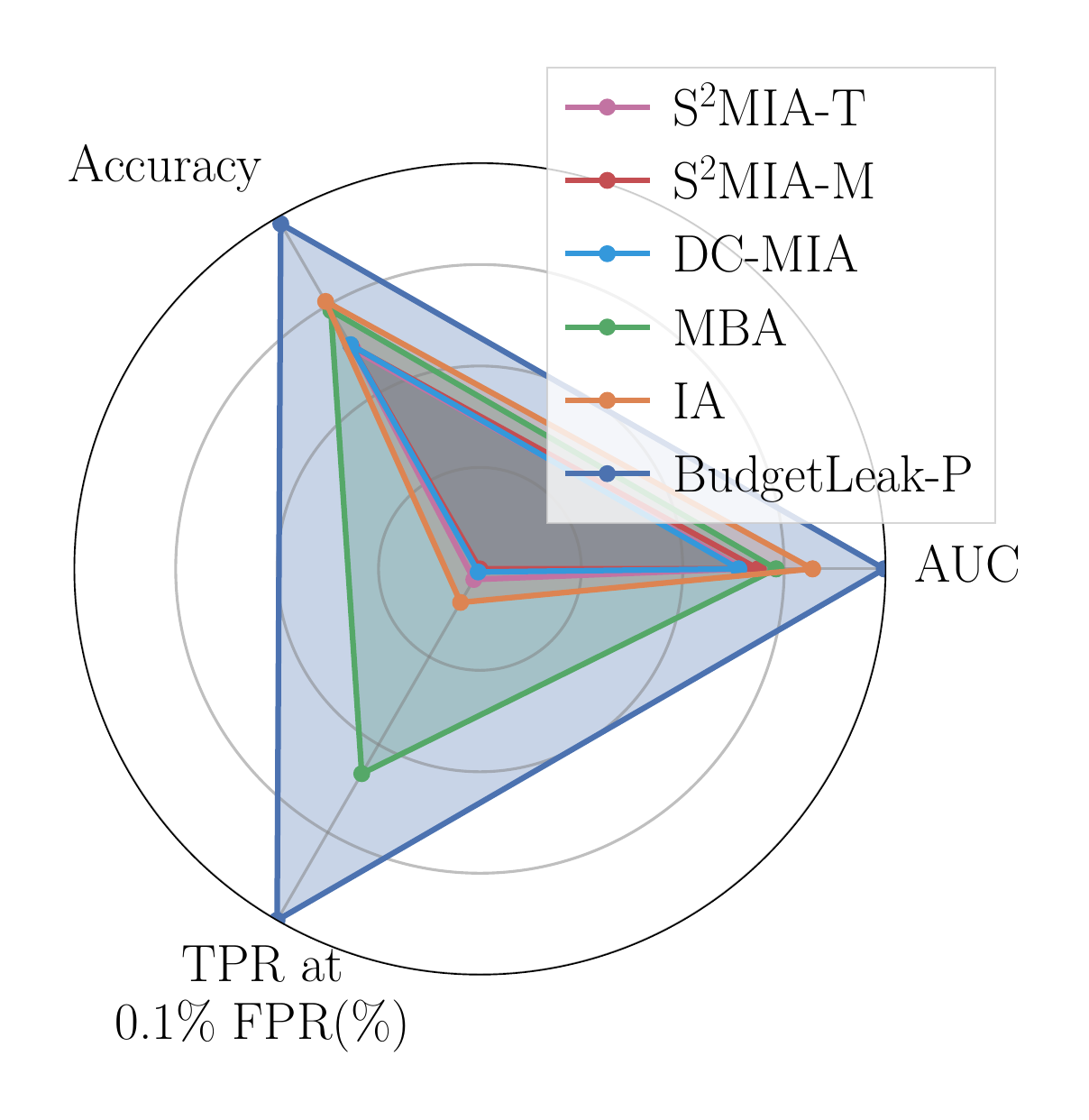}
        \caption{MiniLM}
        \label{fig:miniLM}
    \end{subfigure}
    \hfill
    \begin{subfigure}[b]{0.3\linewidth}
        \centering
        \includegraphics[width=\linewidth]{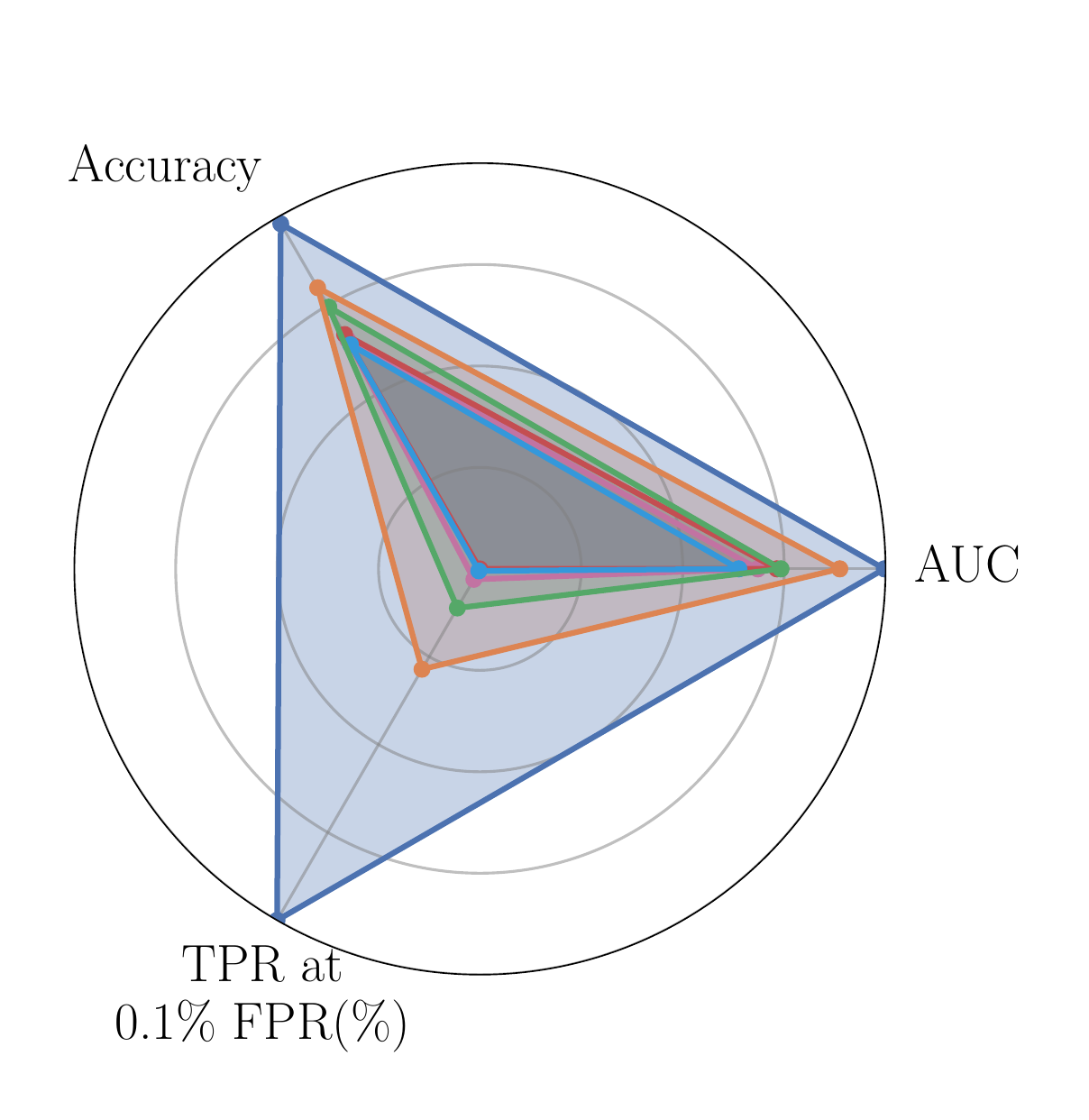}
        \caption{BGE}
        \label{fig:bge}
    \end{subfigure}
    \hfill
    \begin{subfigure}[b]{0.3\linewidth}
        \centering
        \includegraphics[width=\linewidth]{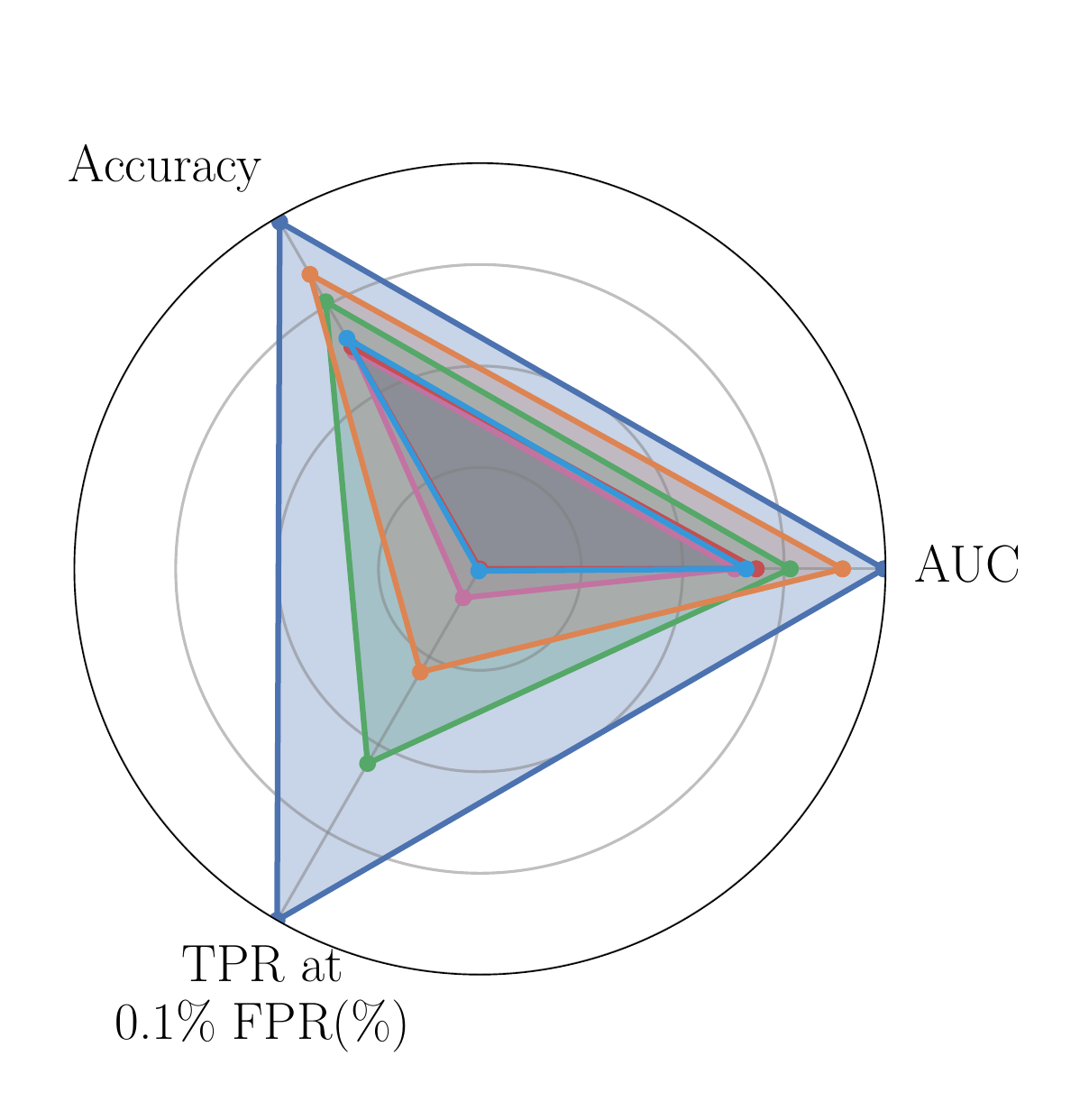}
        \caption{Ideal}
        \label{fig:ideal}
    \end{subfigure}
    \caption{Attack performance on RAGs built on LLaMA with different retrievers on the HealthCareMagic-100k dataset.}
    \label{performance:Health_LLaMA_retriver}
\end{figure*}


\begin{figure*}[h]
    \centering
    \begin{subfigure}[b]{0.3\linewidth}
        \centering
        \includegraphics[width=\linewidth]{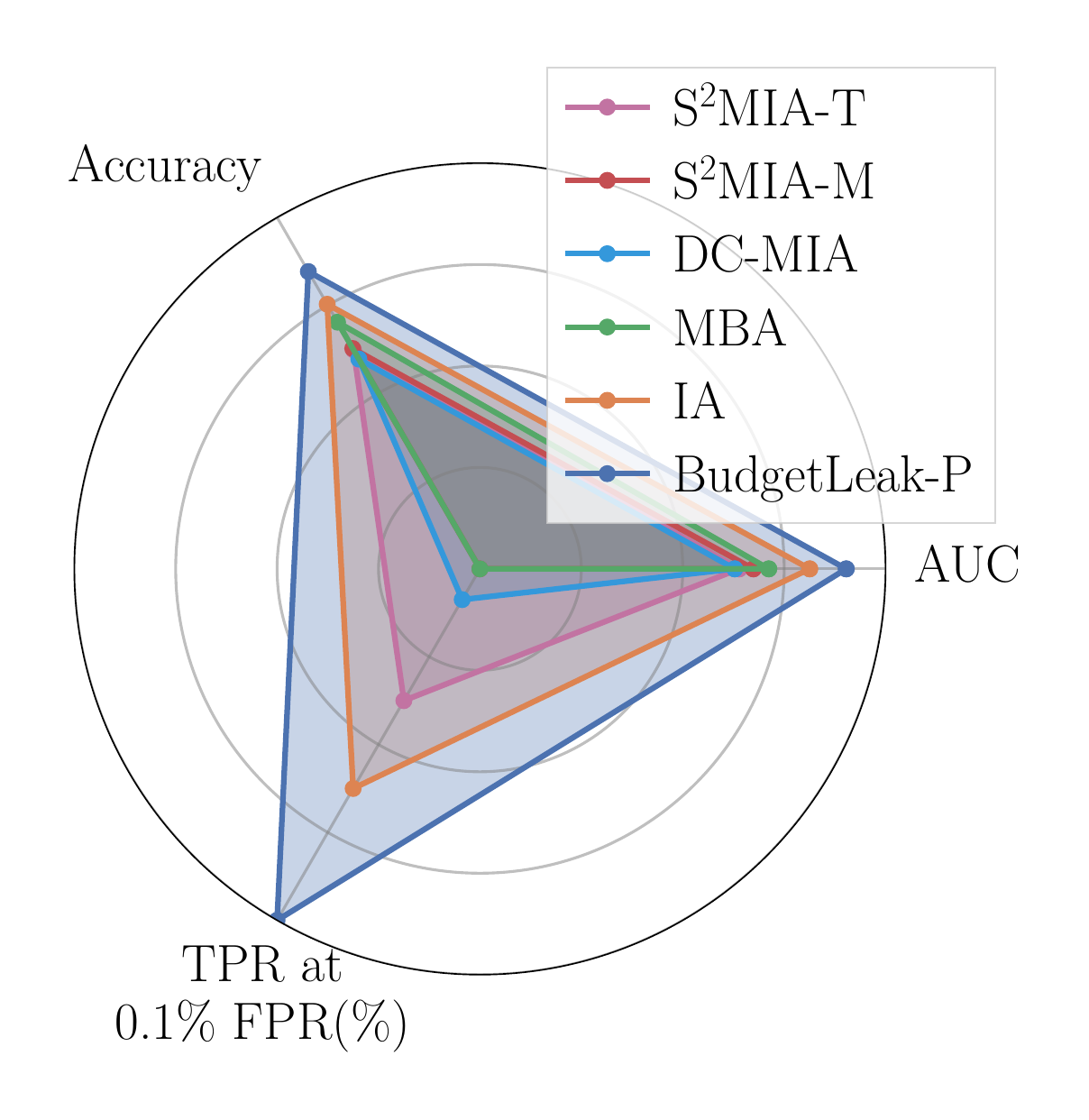}
        \caption{MiniLM}
        \label{fig:NQ_miniLM}
    \end{subfigure}
    \hfill
    \begin{subfigure}[b]{0.3\linewidth}
        \centering
        \includegraphics[width=\linewidth]{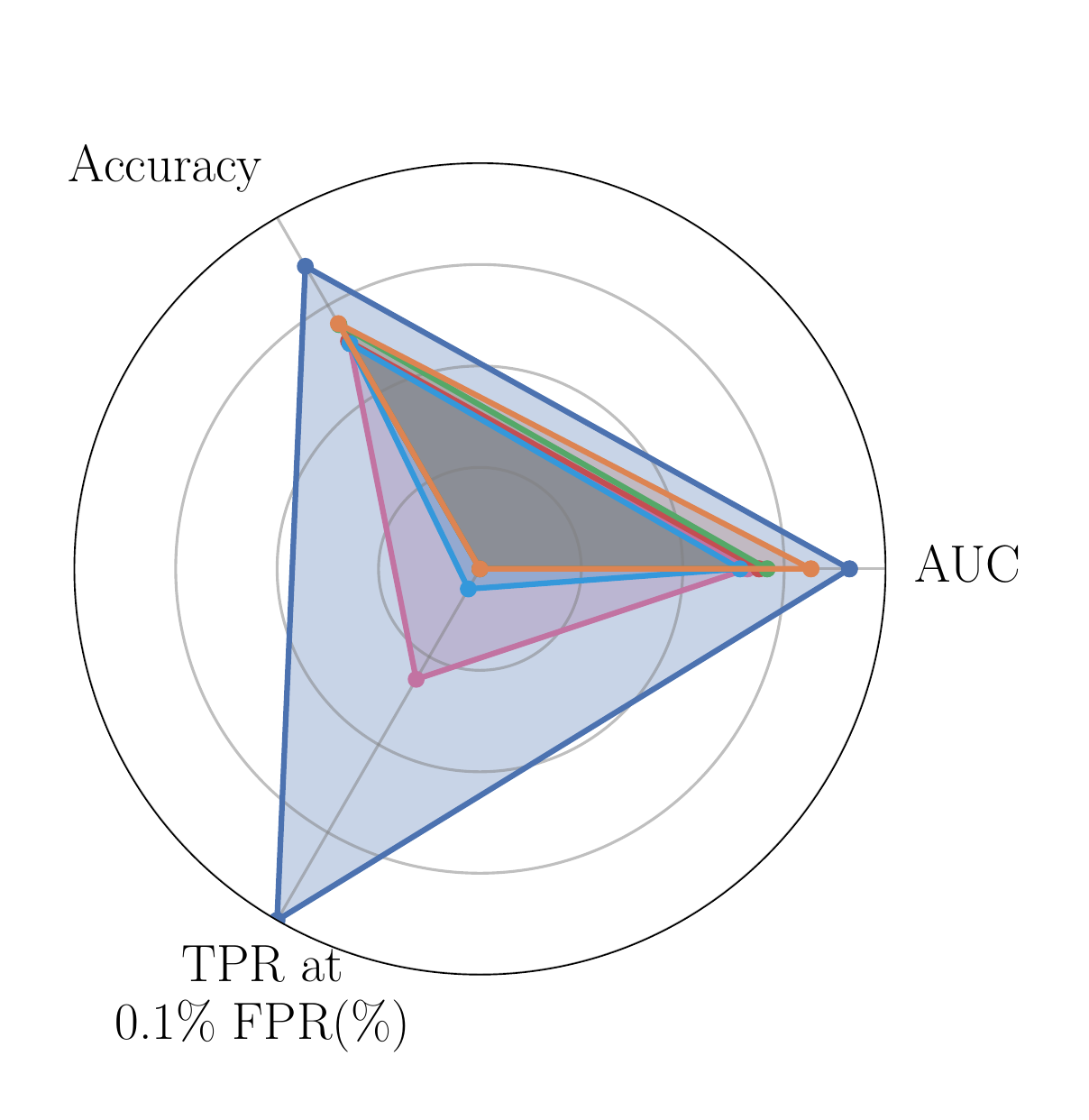}
        \caption{BGE}
        \label{fig:NQ_bge}
    \end{subfigure}
    \hfill
    \begin{subfigure}[b]{0.3\linewidth}
        \centering
        \includegraphics[width=\linewidth]{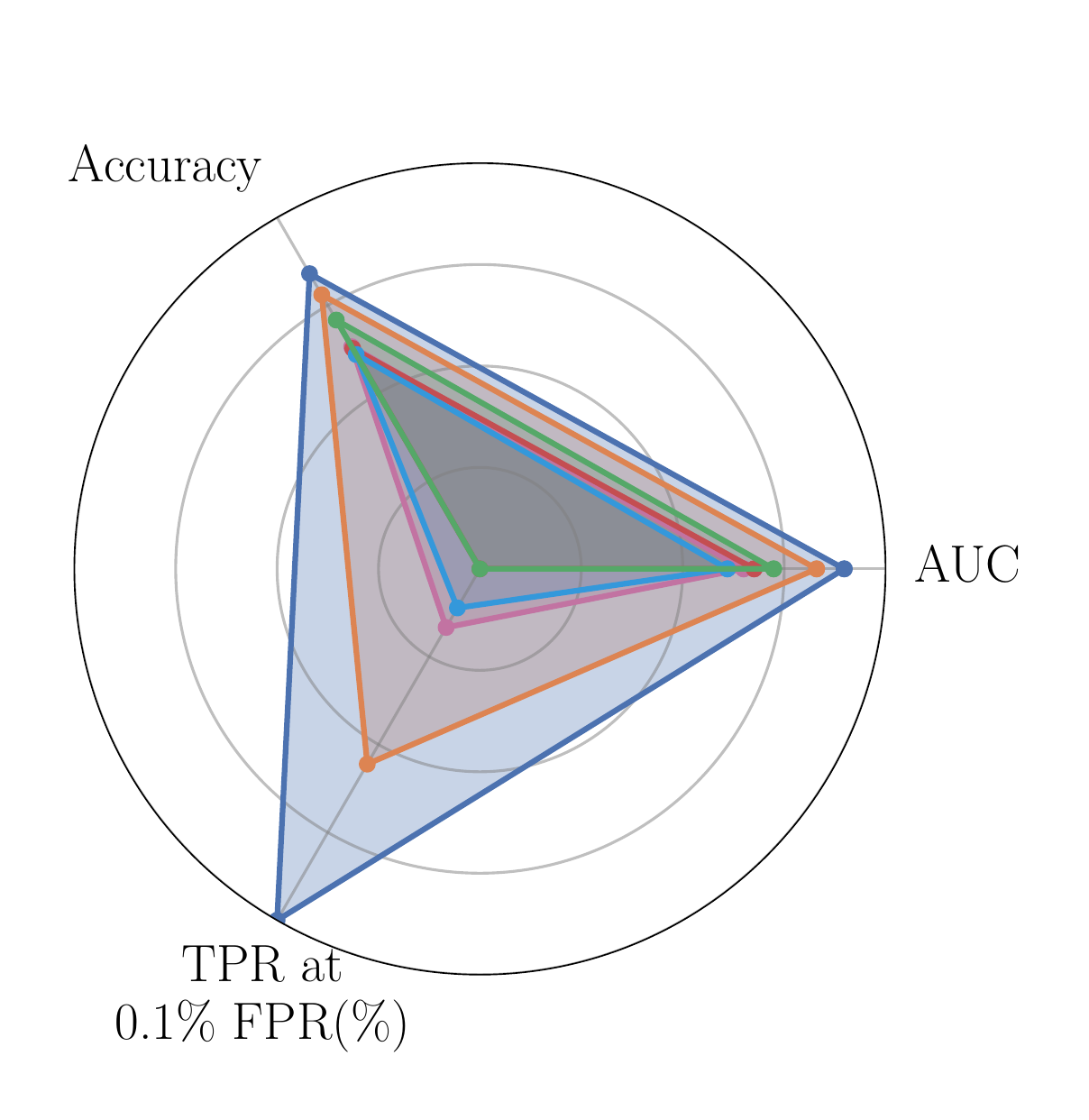}
        \caption{Ideal}
        \label{fig:NQ_ideal}
    \end{subfigure}
    
    \caption{Attack performance on RAGs built on LLaMA with different retrievers on the Natural Questions dataset.}
    \label{performance:NQ_LLaMA_retriver}
\end{figure*}


\subsection{Overall Performance}\label{overallperformance}
\mypara{Performance of BudgetLeak-P} \autoref{performance:healthcaremagic}, \autoref{performance:MSMarco}, \autoref{performance:NQ}, and \autoref{performance:agnews} compare the performance of BudgetLeak-P with baseline attacks. The target RAG in these experiments uses MiniLM as the retriever, and the impact of different retrievers is further analyzed in \autoref{Ablation-Study}. 

BudgetLeak-P consistently achieves the best results across all cases. In particular, for TPR at 0.1\% FPR, it outperforms baselines by a large margin. For example, in \autoref{performance:MSMarco}, when attacking a RAG system with LLaMA on MS-MARCO, BudgetLeak-P achieves a TPR of 11.5\%, while the best baseline reaches only 1.5\%, representing a nearly 8× improvement. In the same setting, BudgetLeak-P attains an accuracy of 0.959, compared to 0.734 from the best baseline—an absolute gain of 0.225, or a 30\% relative improvement. 

In contrast, baseline attacks exhibit unstable performance across datasets. For instance, IA achieves the second-best AUC against ChatGLM on HealthCareMagic-100k (\autoref{performance:healthcaremagic}), yet ranks among the worst when attacking ChatGLM on MS-MARCO (\autoref{performance:MSMarco}).

The superior and stable performance of BudgetLeak-P stems from its ability to exploit a previously overlooked generation-budget side channel, where output quality strongly correlates with output length. Consequently, member queries exhibit significant variations in response quality under different output length constraints, while non-member queries show less pronounced changes. These differences provide clear membership signals that BudgetLeak-P effectively leverages, leading to consistently strong performance across all evaluated datasets.


\mypara{Performance of BudgetLeak-Z}
Here, we consider a more realistic and challenging attack setting where the adversary has no internal knowledge of the target RAG system and no access to data from the same distribution as its knowledge base. While some baseline methods were not designed for this scenario, we adapt them for comparison by applying the same clustering strategy as our BudgetLeak-Z. Specifically, we use semantic similarity for S$^2$MIA, fill-in accuracy for MBA, response accuracy for IA, and calibrated score for DC-MIA, followed by FCM clustering to assign each sample a fuzzy membership score. These scores are then used to compute TPR at 0.1\% FPR and AUC. For accuracy, we classify samples with a membership score above 0.5 as members.

\autoref{cluster} compares the performance of BudgetLeak-Z with baseline methods. In most cases, BudgetLeak-Z achieves the best or second-best results. For instance, when attacking LLaMA with HealthCareMagic-100k, BudgetLeak-Z attains an AUC of 0.978, while the strongest baseline reaches only 0.730, which corresponds to a 34\% relative improvement. This highlights the strong discriminative power of the membership signal derived from the generation budget side channel. Although BudgetLeak-Z does not always outperform all baselines, its performance remains consistently close to the best. Notably, applying fuzzy clustering to the baseline membership scores is an integral part of our methodology, which was not considered in the original designs of these baselines.

\subsection{Ablation Study}\label{Ablation-Study}
In this section, we investigate several factors that influence the effectiveness of BudgetLeak attacks.

\mypara{Retrievers} The retriever is central to RAG systems. To assess its impact on attack performance, we test our method and baselines with an alternative embedding model, BGE~\cite{bge-small-en-v1.5}, in addition to the default MiniLM~\cite{all-MiniLM-L6-v2} used in \autoref{overallperformance}. We also design an ideal retriever that always returns the target member sample at the top of the context, serving as an upper bound on retrieval quality. As shown in \autoref{performance:Health_LLaMA_retriver} and \autoref{performance:NQ_LLaMA_retriver}, BudgetLeak-P consistently outperforms all baselines across retrievers and settings. 
At a 0.1\% FPR, it achieves a markedly higher TPR than competing methods. Additional results on BGE are provided in Appendix~\autoref{bge_mistral}.

This robustness stems from two factors. First, we use the question part of each sample, which reliably retrieves the correct QA pair. Second, our side-channel manipulation amplifies the membership signal, making it easier to detect.



\begin{figure*}[!t]
\centering
\begin{subfigure}{0.6\columnwidth}
\includegraphics[width=\columnwidth]{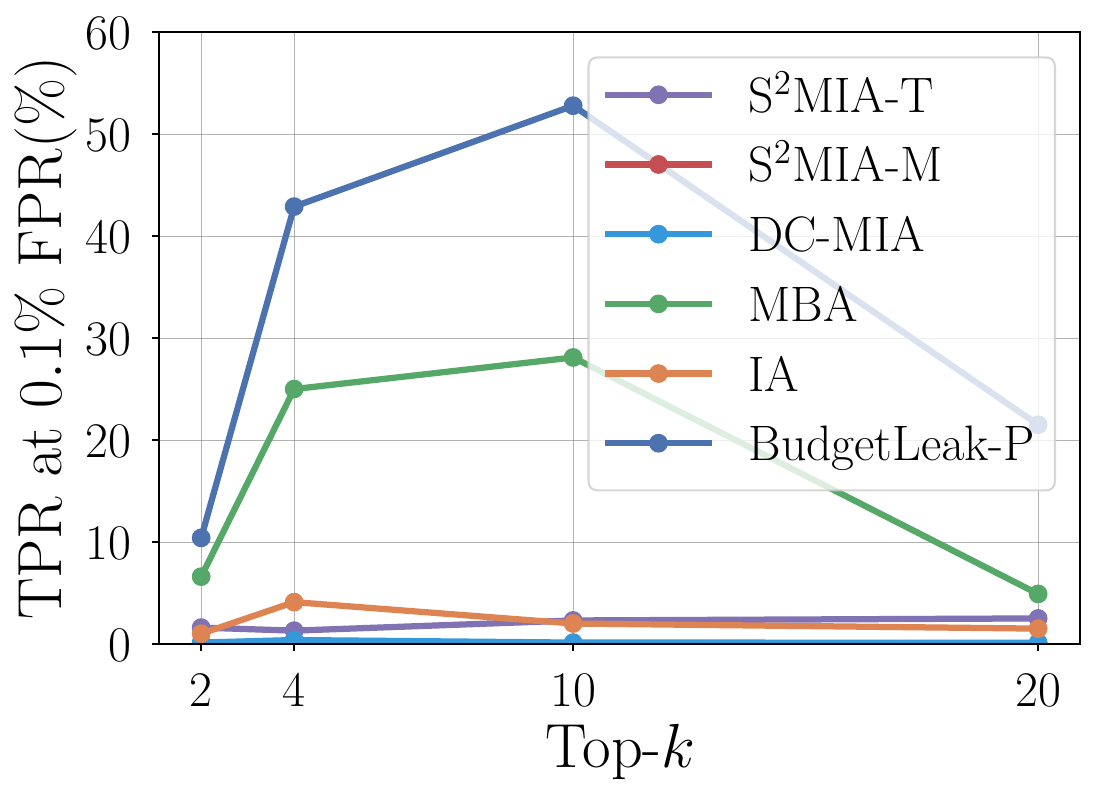}
\label{fig:topk_tpr}
\end{subfigure}
\hspace{20pt} 
\begin{subfigure}{0.6\columnwidth}
\includegraphics[width=\columnwidth]{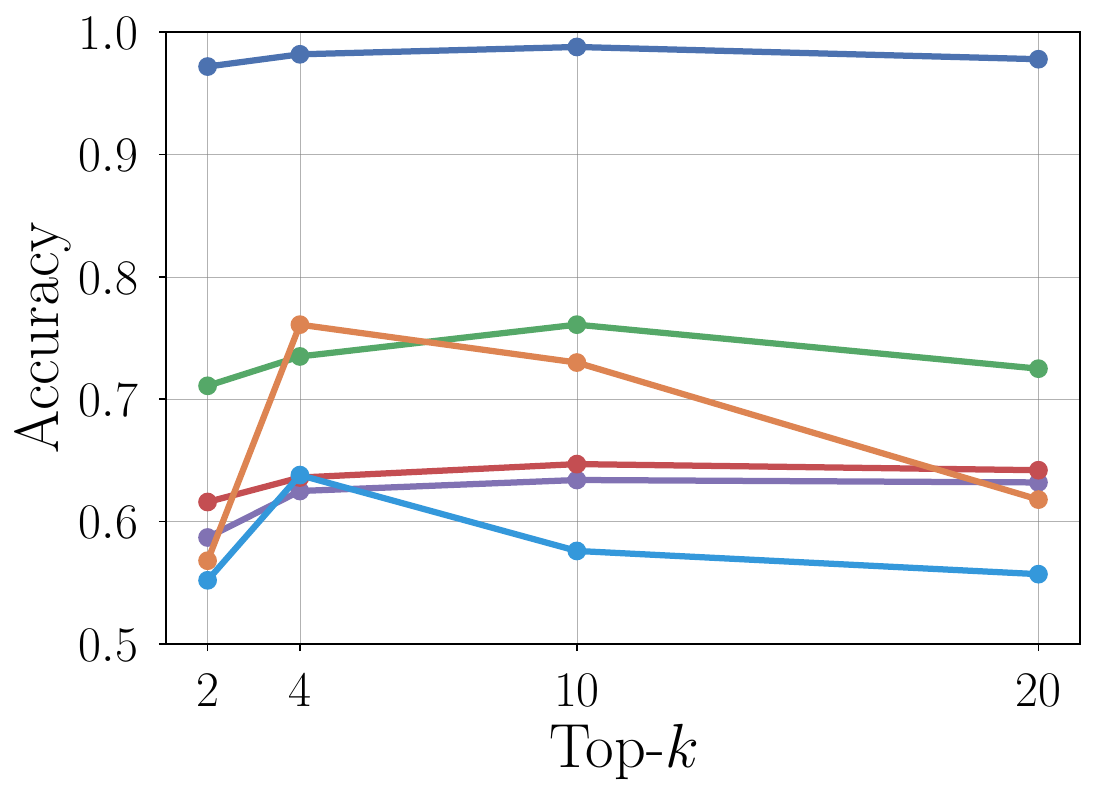}
\label{fig:topk_acc}
\end{subfigure}
\hspace{20pt} 
\begin{subfigure}{0.6\columnwidth}
\includegraphics[width=\columnwidth]{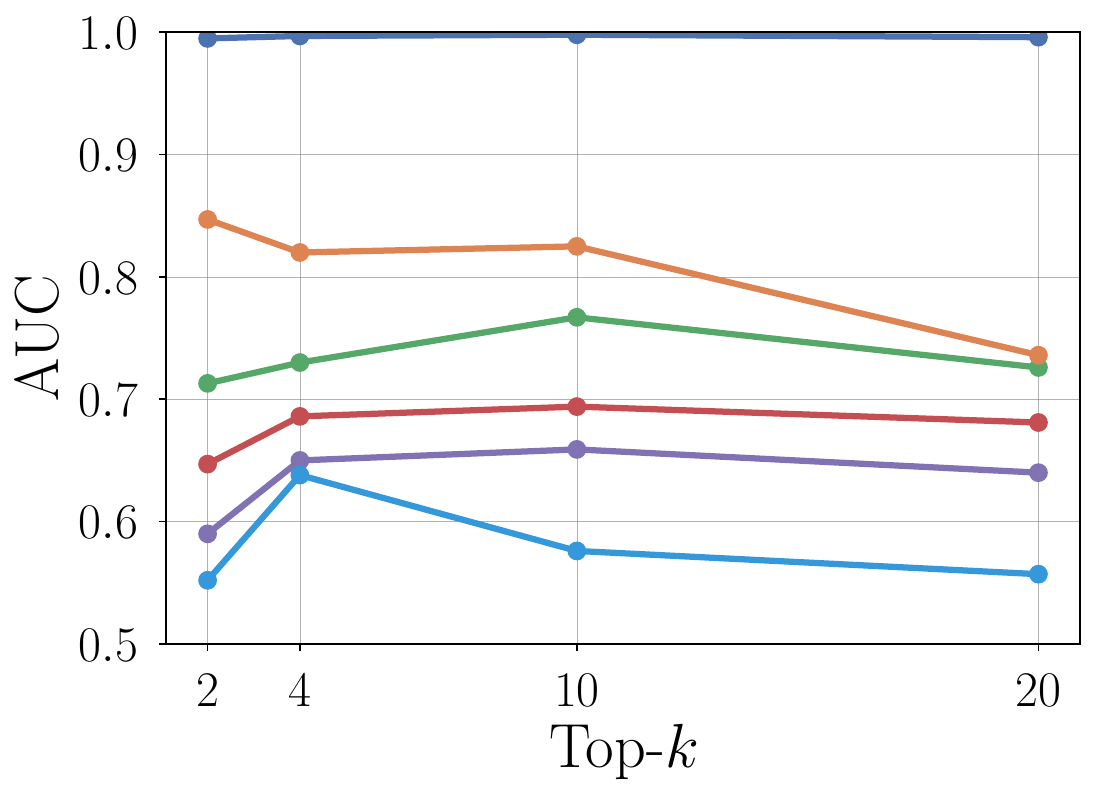}
\label{fig:topk_auc}
\end{subfigure}

\caption{Impact of retrieved item count (top-$k$) on attack performance on RAG with LLaMA over HealthCareMagic-100k.}
\label{topk_results}
\end{figure*}

\mypara{Number of Retrieved Items (top-$k$)}
Generally, a larger top-$k$ increases the likelihood of retrieving the target sample but also leads to longer context lengths, thereby imposing a greater computational burden on the RAG system. Therefore, we evaluate the impact of top-$k$ on attack performance. As shown in \autoref{topk_results}, for most attack methods, including BudgetLeak-P, initially increasing top-$k$ improves attack effectiveness due to a higher probability of retrieving the target sample. However, as top-$k$ continues to grow, the attack performance tends to decline. This is because the context becomes cluttered with many less relevant items, which degrades the quality of RAG’s responses and weakens the distinction between members and non-members. Notably, throughout the increase of top-$k$, BudgetLeak consistently outperforms other baselines, demonstrating its robustness. For example, when top-$k$ increases from 2 to 20, BudgetLeak-P maintains an accuracy above 0.972, whereas all other baselines remain below 0.761.

\mypara{Different Metric Sequences}
In the overall performance comparison shown in \autoref{overallperformance}, we employ a range of evaluation metrics, including semantic similarity (Cosine Similarity) and lexical metrics such as ROUGE (ROUGE-1, ROUGE-2, ROUGE-L), BLEU, and Edit Distance. This section further investigates which types of metrics carry stronger membership signals and whether it is necessary to use all of them. For simplicity, we select one representative metric from each category. As shown in \autoref{different_metric_sequences}, lexical metrics generally capture stronger membership information. For example, using only the BLEU sequence, BudgetLeak-P achieves the second-best performance in terms of AUC, Accuracy, and TPR at 0.1\% FPR on the Natural Questions dataset. In contrast, semantic similarity tends to be less effective when used alone. However, combining both semantic and lexical metrics leads to the best overall results, indicating that these metrics provide complementary membership signals. We recommend using the full set when computational resources allow, and prioritizing lexical metrics in resource-constrained scenarios.

\begin{table}[!htbp]
\caption{Impact of metric sequence variants on attack performance on RAG with LLaMA.}
\label{different_metric_sequences}
\centering
\resizebox{\linewidth}{!}{ 
\setlength{\tabcolsep}{4pt}
\renewcommand{\arraystretch}{1}
\begin{tabular}{@{}c|cccc@{}}
\toprule
Dataset & Metric & AUC & Accuracy & \makecell{TPR at 0.1\% \\ FPR(\%)} \\ 
\midrule
\multirow{5}{*}{HealthCareMagic-100k}
 & Cosine Similarity & 0.794	& 0.736	& 1.6
 \\  
 & BLEU              & \underline{0.987}	& 0.947	& 21.6
 \\ 
 & ROUGE-1           & 0.985	& 0.955	& 24.5
 \\ 
 & Edit Distance     & 0.985	& \underline{0.958}	& \underline{42.0}
 \\  
 & Multi-metric      & \textbf{0.997}	& \textbf{0.982}	& \textbf{42.9}
 \\  
\midrule
\multirow{5}{*}{Natural Questions} 
 & Cosine Similarity & 0.749	& 0.696	& 0.6
 \\  
 & BLEU              & \underline{0.867}	& \underline{0.817}	& \underline{1.0}  
 \\ 
 & ROUGE-1           & 0.850 & 0.783	& \underline{1.0}
 \\ 
 & Edit Distance     & 0.834	& 0.780 & 0.2
 \\  
 & Multi-metric      & \textbf{0.903}	& \textbf{0.846}	& \textbf{1.6}
 \\  
\bottomrule
\end{tabular}
}
\end{table}

\begin{table*}[!htbp]									
\caption{Impact of attack model architecture on attack performance across multiple RAG systems.}										
\label{Attack_Model}	
\centering												
\resizebox{\textwidth}{!}{			
\setlength{\tabcolsep}{8pt}		
\renewcommand{\arraystretch}{1}			
\begin{tabular}{@{}c|cccc|cccc@{}}					
\toprule												
\multirow{2}{*}{Dataset} & \multicolumn{4}{c|}{LLaMA} & \multicolumn{4}{c}{Mistral} \\												
\cmidrule(lr){2-5} \cmidrule(lr){6-9}	
 & Method & AUC & Accuracy & \makecell{TPR at 0.1\% \\ FPR(\%)} & Method & AUC & Accuracy & \makecell{TPR at 0.1\% \\ FPR(\%)} \\	
\midrule												
\multirow{5}{*}{HealthCareMagic-100k}				
 & LSTM                  & 0.991	& 0.980	& \underline{36.8}						
 & LSTM                  & \underline{0.882}	& \underline{0.811}	& \underline{14.4}\\ 					
 & Attention-based LSTM  & \textbf{0.997}	& \underline{0.982}	& \textbf{42.9}
 & Attention-based LSTM  & \textbf{0.892}	& \textbf{0.812}	& \textbf{14.6}\\							
 & Transformers          & 0.959	& 0.859	& 32.4								
 & Transformers          & 0.840	& 0.628	& 6.6\\ 
 & Mamba                 & \underline{0.996}	& \textbf{0.983}	& 30.2			
 & Mamba                 & 0.881	& 0.805	& \underline{14.4}\\ 
\midrule												
\multirow{5}{*}{MS-MARCO}
 & LSTM                  & 0.979	& 0.953	& \underline{11.4}						
 & LSTM                  & \underline{0.952}	& 0.893	& \underline{9.9}\\ 					
 & Attention-based LSTM  & \underline{0.980} & \underline{0.959}	& \textbf{11.5}
 & Attention-based LSTM  & \textbf{0.953}	& \textbf{0.896}	& \textbf{17.4} \\							
 & Transformers          & \underline{0.980} & 0.953	& 4.5								
 & Transformers          & 0.936	& 0.779	& 0.6 \\ 
 & Mamba                 & \textbf{0.986}	& \textbf{0.967}	& 5.3			
 & Mamba                 & 0.943	& \underline{0.895}	& 2.7\\ 
\bottomrule												
\end{tabular}												
}												
\end{table*}


\mypara{Architecture of Attack Model}
Our method hinges on exploiting the differing sensitivities of RAG systems’ response quality to generation budget variations between member and non-member samples for membership inference. For BudgetLeak-P, we construct multi-metric sequences and use sequence-based attack models to capture these distinct behavioral patterns. We compare several architectures including LSTM~\cite{lstm}, Attention-based LSTM~\cite{atae_lstm,li2024tulam}, Transformers~\cite{transformer}, and the recently proposed Mamba~\cite{gu2023mamba}. As shown in \autoref{Attack_Model}, Attention-based LSTM typically achieves the best or second-best results, while Transformers and Mamba underperform, likely due to the short length of the sequences, which limits their advantage in modeling long dependencies. Therefore, we adopt Attention-based LSTM as our default attack model.

\section{Discussion}\label{discussion}
In this section, we discuss the robustness, practicality, and limitations of BudgetLeak.

\subsection{Evaluation of Robustness}

To mitigate jailbreak attacks~\cite{liu2023autodan,jia2024improved}, LLM-based systems, including RAG, often apply query rewriting to preserve semantics while altering surface form~\cite{wang2025maferw,ma2023query}. This disrupts MIAs that rely on carefully crafted inputs. Inspired by defenses such as AdvReg~\cite{nasr2018machine}, which perturb output probabilities in classification models, we also consider response rewriting in the RAG setting. In this section, we evaluate how such rewriting techniques affect MIA performance.

Following~\cite{naseh2025riddle}, we use GPT-4o~\cite{achiam2023gpt} to paraphrase both queries and responses via a rewriting prompt. Results are summarized in \autoref{rewriting} and Appendix \autoref{rewriting-mistral}.

BudgetLeak-P remains the most effective attack under both query and response rewriting defenses, consistently outperforming all baselines. For instance, in \autoref{rewriting}, under response rewriting, BudgetLeak-P achieves an AUC of 0.991, compared to 0.636 from the best baseline. Its robustness stems from leveraging generation-budget side channels by modeling semantic and non-semantic quality dynamics, rather than depending on static output content, making it less sensitive to rewriting.

Among the two defenses, response rewriting proves more effective. By altering the output of the model, rewriting with an LLM distorts observable membership cues. For instance, fill-in accuracy used by MBA drops significantly when masked tokens are paraphrased with synonyms, reducing discriminability. As shown in \autoref{rewriting}, MBA’s AUC falls to 0.584 under response rewriting. IA is even more affected, with performance degrading to near-random levels. IA formulates 30 Yes/No questions per sample and infers membership based on answer accuracy. However, rewritten responses often substitute Yes/No/Unknown with expressions like Affirmative, Not specified, or Negative, breaking alignment with IA's scoring logic. While one could add an output polarity classifier to recover original intent, even with perfect recovery, IA still underperforms BudgetLeak-P. For example, IA (no defense) achieves an AUC of 0.820, whereas BudgetLeak-P under response rewriting yields a higher AUC of 0.991 (see \autoref{rewriting}).

In summary, our BudgetLeak-P demonstrates strong robustness against both rewriting-based defenses by exploiting the generation budget side channel, consistently outperforming prior attacks.

\begin{table}[!htbp]
\caption{Attack performance on RAG with LLaMA over HealthCareMagic-100k under query or response rewriting. Reductions relative to the no-rewriting setting are shown in parentheses.}
\label{rewriting}
\centering
\resizebox{\linewidth}{!}{ 
\setlength{\tabcolsep}{4pt}
\renewcommand{\arraystretch}{1.0}
\begin{tabular}{@{}c|ccc@{}}
\toprule
Defense & Method & AUC & Accuracy \\
\midrule
\multirow{5}{*}{Query Rewriting}
 & S$^2$MIA-T & 0.594 (0.056) & 0.595 (0.030)  \\  
 & S$^2$MIA-M & 0.656 (0.030) & 0.611 (0.025)  \\ 
 & DC-MIA     & 0.533 (0.037) & 0.533 (0.037)  \\ 
 & MBA        & 0.647 (0.083) & 0.637 (0.098)  \\ 
 & IA         & \underline{0.818} (0.002) & \underline{0.756} (0.005)  \\ 
\cline{2-4}
 \addlinespace[1ex] 
 & BudgetLeak-P     & \textbf{0.984} (0.013) & \textbf{0.970} (0.012)  \\ 
\midrule
\multirow{5}{*}{Response Rewriting} 
 & S$^2$MIA-T & 0.603 (0.047) & 0.593 (0.032)  \\ 
 & S$^2$MIA-M & \underline{0.636} (0.050) & \underline{0.605} (0.031)  \\
 & DC-MIA     & 0.561 (0.009) & 0.561 (0.009)  \\ 
 & MBA        & 0.584 (0.146) & 0.582 (0.153)  \\ 
 & IA         & 0.490 (0.330) & 0.493 (0.268)  \\
 \cline{2-4}
 \addlinespace[1ex] 
 & BudgetLeak-P     & \textbf{0.991} (0.006) & \textbf{0.962} (0.020)  \\
\bottomrule
\end{tabular}
}
\end{table}

\subsection{Practical Investigation}\label{practialdiscussion}
To observe the trend in answer quality under varying generation budgets, BudgetLeak performs multiple queries to the RAG system using different budget limits. By default, we adopt the Budget Sweep Strategy, issuing queries with generation budgets ranging from 10 to 270 tokens in steps of 20. While effective, this approach incurs high query costs (due to per-token charges) and may increase the risk of detection. Thus, we explore whether the number of queries can be reduced without sacrificing attack effectiveness.

Specifically, we evaluate a Tri-Budget Setting, using three representative generation budgets: a minimal budget of 10 tokens (Min), the average token length in the dataset (Mean), and twice the mean value (Max). These values correspond to compact, economical, and relaxed RAG response conditions, respectively.

As shown in \autoref{practical_investigation_msmarco}, the Tri-Budget setting often achieves AUC and accuracy comparable to the full sweep approach, though it exhibits a notable decrease in TPR at 0.1\% FPR. For instance, on a RAG system with LLaMA, the Tri-Budget setting attains an AUC of 0.974 and an accuracy of 0.940, slightly below the Sweep’s 0.980 AUC and 0.959 accuracy, while the TPR at 0.1\% FPR drops more substantially from 11.5 to 1.3. Even with only two queries, as in the Bi-Budget(MeanMax) setting, the attack still achieves strong results with an AUC of 0.974 and accuracy of 0.939, only marginally lower than the Tri-Budget configuration. This demonstrates that BudgetLeak can generate robust membership inference signals with just two or three queries.

Moreover, BudgetLeak with the Tri-Budget setting consistently outperforms all baselines in AUC and accuracy while maintaining comparable TPR at 0.1\% FPR. For example, under the MS-MARCO and LLaMA setup, the Tri-Budget setting achieves an AUC of 0.974, accuracy of 0.940, and TPR at 0.1\% FPR of 1.3. In contrast, while S$^2$MIA-M attains the highest AUC among baselines in this case, it only reaches 0.789 AUC, 0.731 accuracy, and a TPR of 0 at 0.1\% FPR. Additionally, a recent method, IA, requires 30 queries per target sample to attain its optimal performance. In contrast, BudgetLeak achieves superior results using only three queries, demonstrating significant efficiency and practical advantages.

\begin{table}[!htbp]
\caption{Attack performance under different query budget settings on RAG with MS-MARCO.}
\label{practical_investigation_msmarco}
\centering
\resizebox{\linewidth}{!}{ 
\setlength{\tabcolsep}{4pt}
\renewcommand{\arraystretch}{1}
\begin{tabular}{@{}c|cccc@{}}
\toprule
Model & Method & AUC & Accuracy & \makecell{TPR at 0.1\% \\ FPR(\%)} \\ 
\midrule
\multirow{6}{*}{LLaMA}
 & Bi-Budget(MinMean) & 0.960 & 0.906 & 1.1

 \\  
 & Bi-Budget(MinMax) & 0.960 & 0.918	& \underline{2.5}

 \\ 
 & Bi-Budget(MeanMax) & \underline{0.974}	& 0.939	& 1.1

 \\ 
 & Tri-Budget & \underline{0.974} & \underline{0.940} & 1.3

 \\
 & Sweep & \textbf{0.980}	& \textbf{0.959}	& \textbf{11.5}

  \\
\midrule
\multirow{6}{*}{Mistral} 
 & Bi-Budget(MinMean) & 0.942 & 0.884 & 1.4
 \\  
 & Bi-Budget(MinMax) & \underline{0.955}	& \underline{0.901}	& 3.9
 \\ 
 & Bi-Budget(MeanMax) & 0.942 & 0.892	& \underline{8.4}
 \\ 
 & Tri-Budget & \textbf{0.956} & \textbf{0.902} & 7.6
 \\
 & Sweep & 0.953	& 0.896	& \textbf{17.4}
  \\ 
\bottomrule
\end{tabular}
}
\end{table}

\subsection{Limitations}
BudgetLeak is designed for traditional RAG systems, especially those in question answering settings where responses are grounded in specific retrieved passages. In such contexts, the goal is to infer whether a particular text segment exists in the underlying knowledge corpus. However, our method is not directly applicable to emerging RAG variants such as Graph-RAG~\cite{edge2024local,han2024retrieval,peng2024graph}, where responses are influenced by global knowledge aggregation through structured knowledge graphs. These systems rely less on isolated passage retrieval and more on reasoning over interconnected entities and relations, which complicates the extraction of localized membership signals. Extending membership inference techniques to such globally-aware RAG architectures remains an open and important direction for future research.

\section{Conclusion}
In this work, we identify a previously unexplored side channel in RAG systems: the generation budget, which specifies the maximum token limit for responses. We show that varying this parameter leads to different patterns in output quality between member and non-member queries, particularly in semantic and lexical similarity to the ground-truth answers. Leveraging this insight, we propose BudgetLeak, a membership inference framework that issues multiple queries with different generation budgets and analyzes the resulting fidelity metrics. BudgetLeak includes both a supervised variant using a sequential model (i.e., BudgetLeak-P) and an unsupervised version based on clustering (i.e., BudgetLeak-Z). Extensive experiments across diverse datasets, retrievers, and generators demonstrate that BudgetLeak consistently outperforms existing baselines. Our findings highlight the generation budget as a practical and robust source of membership leakage in RAG systems, emphasizing the need for improved data protection in this growing application domain.





%

\bibliographystyle{plain}
\bibliography{sample-base}





\appendices

\section{Additional Results}

\begin{figure}[!th]
    \centering
    \includegraphics[width=0.6\linewidth]{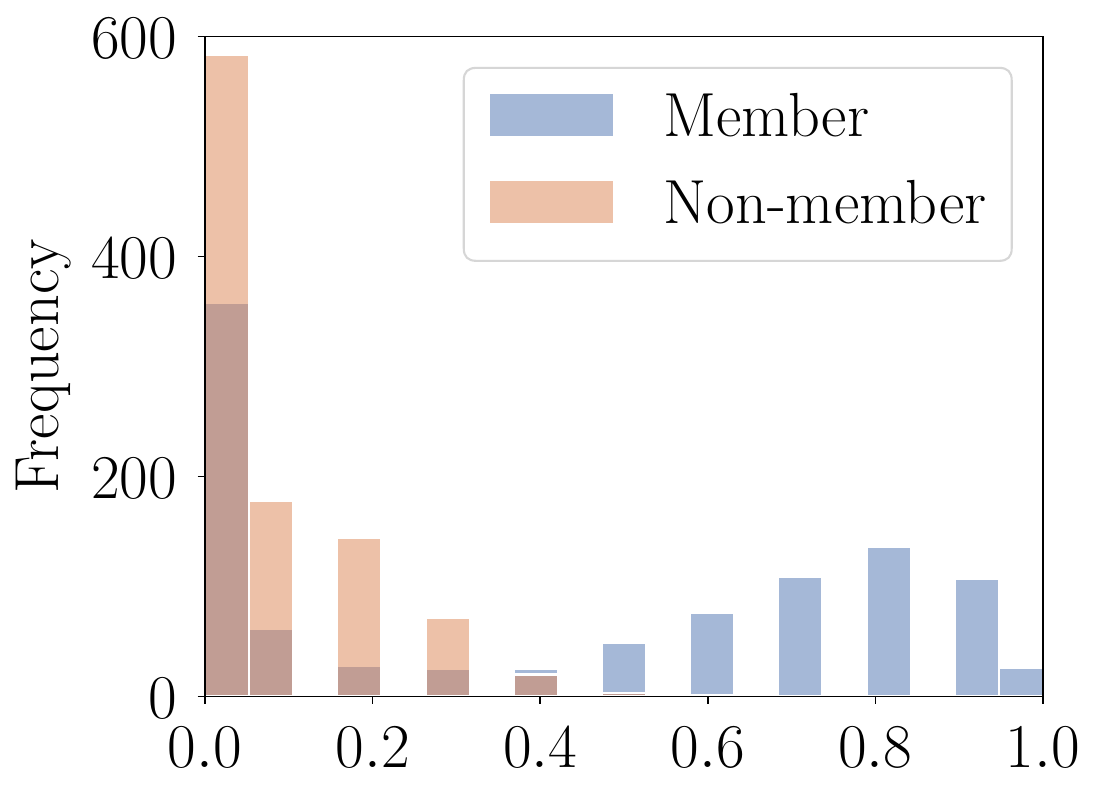}
    \caption{Distribution of fill-in accuracy on RAG with LLaMA over HealthCareMagic-100k.
    }
    \label{fig:fill-inAccuracy}
\end{figure}

\begin{table}[!htbp]
\caption{Attack performance on RAG built on Mistral with BGE as retriever.}
\label{bge_mistral}
\centering
\resizebox{\linewidth}{!}{ 
\setlength{\tabcolsep}{4pt}
\renewcommand{\arraystretch}{1}
\begin{tabular}{@{}c|cccc@{}}
\toprule
Dataset & Method & AUC & Accuracy & \makecell{TPR at 0.1\% \\ FPR(\%)}\\ 
\midrule
\multirow{5}{*}{HealthCareMagic-100k}
& S$^2$MIA-T & 0.712 & 0.654 & 0.0 \\ 
& S$^2$MIA-M & 0.706 & 0.654 & 0.0 \\ 
& DC-MIA & 0.681 & 0.681 & 0.2\\
& MBA & 0.789 & \underline{0.763} & 4.0 \\ 
& IA & \underline{0.851} & \underline{0.763} & \underline{6.5} \\
\cline{2-5}
\addlinespace[1ex] 
& BudgetLeak-P & \textbf{0.868} & \textbf{0.795} & \textbf{13.4} \\ 
\midrule
\multirow{5}{*}{Natural Questions} 
& S$^2$MIA-T & 0.703 & 0.658 & 1.0 \\ 
& S$^2$MIA-M & 0.698 & 0.658 & 0.0 \\ 
& DC-MIA & 0.772 & \underline{0.772} & 0.3\\
& MBA & 0.737 & 0.703 & 0.0 \\ 
& IA & \underline{0.861} & 0.750 & \underline{1.3} \\ 
\cline{2-5}
\addlinespace[1ex] 
& BudgetLeak-P & \textbf{0.882} & \textbf{0.820} & \textbf{2.9} \\ 
\bottomrule
\end{tabular}
}
\end{table}

\begin{table}[!htbp]
\caption{Attack performance on RAG with Mistral over HealthCareMagic-100k under query or response rewriting. Reductions relative to the no-rewriting setting are shown in parentheses.}
\label{rewriting-mistral}
\centering
\resizebox{\linewidth}{!}{ 
\setlength{\tabcolsep}{4pt}
\renewcommand{\arraystretch}{1.0}
\begin{tabular}{@{}c|ccc@{}}
\toprule
Defense & Method & AUC & Accuracy \\
\midrule
\multirow{5}{*}{Query Rewriting}
 & S$^2$MIA-T & 0.609 (0.052) & 0.579 (0.037)  \\  
 & S$^2$MIA-M & 0.595 (0.049) & 0.574 (0.033)  \\ 
 & DC-MIA     & 0.524 (0.173) &  0.524(0.173)  \\ 
 & MBA        & 0.665 (0.088) & 0.654 (0.073)  \\ 
 & IA         & \underline{0.824} (0.009) & \underline{0.745} (0.023)  \\
 \cline{2-4}
 \addlinespace[1ex] 
 & BudgetLeak-P     & \textbf{0.891} (0.001) & \textbf{0.810} (0.002)  \\ 
\midrule
\multirow{5}{*}{Response Rewriting} 
 & S$^2$MIA-T & 0.644 (0.017) & 0.601 (0.015)  \\ 
 & S$^2$MIA-M & 0.641 (0.003) & 0.607 (0.000)  \\
 & DC-MIA     & \underline{0.683} (0.014)  &  \underline{0.683}(0.014)  \\ 
 & MBA        & 0.558 (0.195) & 0.558 (0.169)  \\ 
 & IA         & 0.640 (0.193) & 0.563 (0.205)  \\ 
 \cline{2-4}
 \addlinespace[1ex] 
 & BudgetLeak-P     & \textbf{0.845} (0.047) & \textbf{0.763} (0.049)  \\
\bottomrule
\end{tabular}
}
\end{table}

\end{document}